\documentclass[prb,aps,amsfonts,amssymb,draft,floats,twocolumn]{revtex4}
\usepackage{epsf}

\newcommand{\be}{\begin{equation}}
\newcommand{\ee}{\end{equation}}
\newcommand{\one}{\openone}
\newcommand{\vS}{\mathbf{S}}
\newcommand{\vH}{\mathbf{H}}
\newcommand{\vs}{\mathbf{s}}

\renewcommand{\vr}{\mathbf{r}}
\newcommand{\spacing}{\delta}

\begin{document}

\title{Electron correlations in metal nanoparticles with
spin-orbit scattering}

\author{Denis A. Gorokhov and Piet W. Brouwer}

\affiliation{Laboratory of Atomic and Solid State Physics,
Cornell University, Ithaca, NY 14853-2501, USA}

\begin{abstract}
The combined effect of electron-electron interactions
and spin-orbit scattering
in metal nanoparticles
can be studied by measuring splitting of electron levels in 
magnetic field
($g$ factors) in tunneling spectroscopy
experiments.
Using random matrix theory to describe the single-electron
states in the metal particle, we find that even a relatively small 
electron-electron interaction strength
(ratio of exchange constant $J$ and mean level
spacing $\spacing$  $\simeq 0.3$) significantly
increases $g$-factor fluctuations for
not-too-strong spin-orbit scattering rates (spin-orbit
time $\tau_{\rm so} \gtrsim 1/\spacing$).
In particular, $g$-factors larger than 2 could be observed.
(This is a manifestation of the many-body correlation effects
in nanoparticles). While so far measurements 
only on noble metal (Cu, Ag, Au) and Al samples
have been done for which the effects of
 electron-electron interactions are negligible, 
we discuss   
the possibility of observing 
interaction effects
in nanoparticles made of other metals. 

\end{abstract}

\maketitle

\newpage

\section{Introduction}

While the study of the combined effect of electron-electron
interactions and elastic impurity scattering in two dimensions and
near the metal-insulator transition in three dimensions remains one of
the most important problems in solid state
physics,  the description of electron-electron
interactions in disordered normal metal nanoparticles ({\em i.e.,}
``zero dimensions'') has been found to be remarkably 
simple\cite{kn:kurland2000,kn:aleiner2002}: 

At a 
fixed
number of electrons and without spin-orbit scattering, the only 
relevant interaction term is a long-range exchange
interaction\cite{foot0}
\begin{equation}
  H_{\rm int} = - J \vS^2, \label{eq:HintS}
\end{equation}
that couples to the total spin $\vS$ of the
nanoparticle. The exchange
constant $J$ is closely related to one of the Fermi Liquid
constants of the bulk metal, and is independent of the details of the
impurity configuration inside the nanoparticle. The
interaction Hamiltonian (\ref{eq:HintS}), which is known as
``universal interaction Hamiltonian'', is the only form of the
electron-electron interaction compatible with random matrix
theory.\cite{kn:aleiner2002} 
Random matrix theory provides a valid description 
of single-electron states 
as long as the
dimensionless conductance $g$ of the nanoparticle, which is the ratio
of the Thouless energy $E_{\rm Th}$ and the mean level spacing
$\spacing$, is large.\cite{kn:efetov1983,kn:altshuler1986} 
Residual interaction terms not included in
Eq.\ (\ref{eq:HintS})  are sample specific
and small in comparison to Eq.\ (\ref{eq:HintS}) by, at least, a factor
$1/g$.

In the presence of spin-orbit scattering, spin is also 
randomized, giving rise to both sample-to-sample fluctuations of the
electron-electron interaction and a suppression of the exchange 
interaction (\ref{eq:HintS}). Since the spin-orbit scattering rate
$\gamma_{\rm so} = \hbar/\tau_{\rm so}$ plays the role of a
``Thouless energy'' for the spin degree of freedom, for strong 
spin-orbit scattering the residual exchange interaction becomes 
small by a factor $\gamma_{\rm so}/\spacing\gg 1$ in comparison to the 
interaction strength without spin-orbit scattering if spin-orbit
scattering. However,
the exchange interaction remains the dominant contribution to the
electron-electron interaction as long as $\gamma_{\rm so}/\spacing
\ll g$. 

In this paper, we present a detailed analysis of the combined 
effect of spin-orbit scattering and electron-electron interactions 
in the regime of moderate spin-orbit scattering, $\gamma_{\rm so} 
\sim \spacing$. In this parameter regime, the exchange interaction
is not fully suppressed, while fluctuations are of the same order 
as the average.\cite{kn:adam2002} This makes the
regime of moderate spin-orbit scattering rates qualitatively
different from that without spin-orbit scattering ($\gamma_{\rm so} 
= 0$) and that of strong spin-orbit scattering 
($\gamma_{\rm so}/\spacing \gg 1$.) The parameter regime $\gamma_{\rm
  so} \sim \spacing$ is of interest for recent
experiments on metal 
nanoparticles,\cite{kn:salinas1999,kn:davidovic1999,kn:petta2001,%
kn:petta2002} in which the magnetic-field
dependences of many-electron levels has been measured using
tunneling 
spectroscopy.\cite{kn:ralph1995} Moreover, an analysis of the regime
$\gamma_{\rm so}/\spacing \sim 1$ serves as a model study of the 
breakdown of the ``universal interaction
Hamiltonian'' when the dimensionless conductance is small. A brief
account of some of our findings was previously published in Ref.\
\onlinecite{kn:gorokhov2003}.

Experimentally, the spin structure of electronic states can 
be measured through the magnetic-field dependence of steps
in the current-voltage characteristic of a metal nanoparticle coupled to
source and drain electrodes via tunneling contacts. These steps
occur if the applied voltage is equal to the difference of
the energies of many-electron levels $|N_e+1;k\rangle$ and 
$|N_e;l\rangle$ which have $N_e$ and $N_e+1$ electrons,
respectively,\cite{kn:vondelft2001}
\begin{equation}
  e V_{kl} = E_{N_e+1,k} - E_{N_e,l}.
  \label{eq:V}
\end{equation}
The
derivative $\partial V_{kl}/\partial B$ of the voltage at which a
step occurs to the magnetic field $B$ is parameterized 
through a ``$g$-factor'', 
\begin{equation}
  e \frac{\partial V_{kl}}{\partial B} = \pm \frac{1}{2} g_{kl} \mu_B,
  \label{eq:Vg}
\end{equation}
where $\mu_B = |e| \hbar/2mc$ is the Bohr magneton. 
Without spin-orbit scattering, but with electron-electron 
interactions, many-electron states are characterized by their total 
spin $S$ and by its $z$ component. 
Since tunneling spectroscopy measures transitions in which the
electron number changes by one, the total spin $S$ of the nanoparticle 
changes by $1/2$ upon addition or removal of an electron. This 
``selection rule'' renders all observed $g$ factors equal to two,
irrespective of the spin $S_k$ and $S_l$ of the two many-electron 
levels participating in the transition. 
On the other hand, without interactions, the tunneling spectroscopy
$g$ factors correspond to the magnetic moment of one single-electron
level only, giving rise to a distribution of $g$ factors in which all
levels have, at most, spin 
$1/2$.\cite{kn:brouwer2000,kn:matveev2000,kn:adam2002b}
As was shown in 
Ref.\ \onlinecite{kn:gorokhov2003}, the combined effect of spin-orbit
scattering and electron-electron interactions
is to simultaneously suppress the spin of the many-electron
states and lift the ``selection rules'', causing a much wider distribution
of tunneling spectroscopy $g$ factors than in the non-interacting
case. In particular, there is a significant probability
to find $g$ factors larger than two if spin-orbit scattering is not
too strong ($\gamma_{\rm so}/\spacing \lesssim 2$). The occurrence of
$g$ factors larger than two is a unique signature of the interplay
of electron-electron interactions and spin-orbit scattering.

\begin{figure}[t]
\epsfxsize= 0.8\hsize
\epsffile{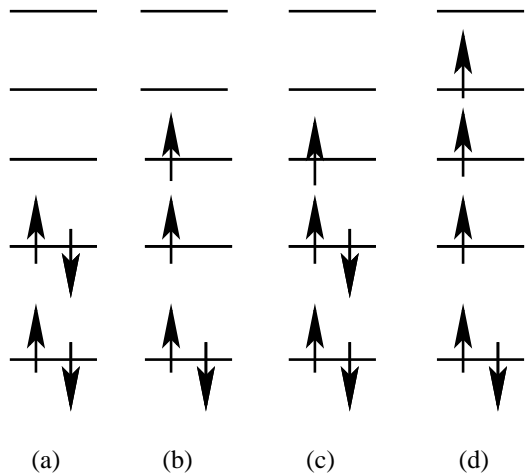}
\caption{\label{fig:1a}
Occupation of single-electron levels for the lowest-lying
many-electron states with total spin $S=0$ (a) and $S=1$ (b),
and for $S=1/2$ (c) and $S=3/2$ (d). In the absence of
exchange interactions, the states with spin $S=0$ and $S=1/2$
are the ground states for even and odd numbers of electrons,
respectively. The exchange interaction compensates (part of) 
the kinetic energy cost of the higher spin states.
}
\end{figure}

What are the main differences between $g$ factor distributions with
and without electron-electron interactions? In order to answer that 
question, we note that
the electron-electron interactions has two main effects in a
nanoparticle: to organize the many-electron 
states according to their total spin $S$, {\em i.e.,} to lift
the degeneracy between states of different $S$ but with the same
orbital content, and to {\em lower} the
energy of a many electron state with spin $S$ by the amount $JS(S+1)$,
increasing the abundance of high-$S$ states among low-energy
excited states, see Fig.\ \ref{fig:1a}. 
(In fact, for $J/\spacing \gtrsim 0.3$, there is a significant
probability that the ground state has a nontrivial spin $S > 1/2$, 
see, {\em e.g.,} 
Refs.\ \onlinecite{kn:brouwer1999b,kn:baranger2000}). It is because
of the combination of these two effects, together with the lifting of 
selection rules by spin-orbit scattering, that electron-electron
interactions enhance the width of the $g$-factor distribution so
significantly. Moreover, because the relative
abundance of high-$S$ states depends on the excitation energy, the
$g$ factor distribution will be different for transitions to an
excited state than for transitions to the ground states. Again, this
is different from the non-interacting case, where 
$g$ factor distributions for transitions to the ground state
and to excited states are equal. 
A third difference between the cases with and without
interactions is that,
as the selection rules are gradually broken down by spin-orbit
scattering, different transitions may have very different weights,
in contrast to the non-interacting case, for which all transitions
have weights within a factor of order unity from each other. (The
``weight'' of the transition is the height of the corresponding
step in the current-voltage characteristics.)

In principle, one should consider contributions to the $g$ factor
from the orbital magnetic moments of the energy levels and from
the spin magnetic moment.\cite{kn:matveev2000} In this work, we 
consider the
spin contribution to the $g$ factor only and neglect the orbital
contribution. For the spin-orbit scattering rates $\gamma_{\rm
  so}/\spacing \sim 1$ we consider here,
this is justified if the electron motion in the metal nanoparticle is
diffusive with mean free path $l$ much smaller than the nanoparticle size
$L$.\cite{kn:matveev2000,kn:adam2002b}

The outline of this paper is as follows. In section II we present
the theoretical formalism. Since we only consider spin-orbit
scattering rates $\gamma_{\rm so}/\spacing \ll g$, random-matrix
theory can be used to describe the single-electron states. 
In Sec.\ III we discuss the results of numerical
simulations for the $g$ factor distribution for transitions from
the $N_e$-electron ground state to the $(N_e + 1)$-electron ground
state and $(N_e+1)$-electron excited states, where $N_e$ is taken
even. The restriction to transitions starting from the $N_e$-electron 
ground state is appropriate if the metal nanoparticle relaxes to the
$N_e$-electron ground state between tunneling events. In Sec.\ IV
we discuss the consequences of our findings for various metals: our
results depend on the ratio $J/\spacing$, which strongly depends on
the metal under consideration. We
conclude in Sec.\ V. Finally, in the appendix, we report an analytical
calculation of the $g$ factor distribution for weak spin-orbit
scattering, $\gamma_{\rm so}/\spacing \ll 1$, again paying special
attention to differences between transitions involving ground states
only and transitions to or from an excited state.

\section{Theoretical Description}
\label{sec:theoretical_description}

\subsection{Effective Hamiltonian}

Random matrix theory can be used to describe the single-electron
wavefunctions and energy levels of a metal nanoparticle. For a metal 
nanoparticle
with spin orbit scattering, the appropriate random matrix ensemble
interpolates between the Gaussian Orthogonal Ensemble (GOE) and the
Gaussian Symplectic Ensemble (GSE) of random matrix theory,
\begin{equation}
  H_0(\lambda) = H_{\rm GOE} +  H_{\rm so}(\lambda ).
  \label{eq:H0}
\end{equation}
Writing the spin degrees of freedom explicitly, one has
\begin{eqnarray}
  H_{\rm GOE} &=& S \otimes \one_2, \\
  H_{\rm so}(\lambda ) &=& 
  \frac{i \lambda}{2 \sqrt{N}}
  \sum_{j=1}^{3} A_j \otimes \sigma_j,
\end{eqnarray}
where $\one_2$ is the $2 \times 2$ unit matrix in spin space,
$\sigma_j$ is the Pauli matrix ($j=1,2,3$), $S$ is an 
$N \times N$ real symmetric matrices, and $A_j$ is a real
antisymmetric matrix ($j=1,2,3$). The elements of the matrices
$S$, $A_1$, $A_2$, and $A_3$ are drawn from independent
Gaussian distributions with zero mean and with equal variances
for the off-diagonal elements. The diagonal elements of $S$
have double variance, whereas the diagonal elements of 
$A_1$, $A_2$, and $A_3$ are zero because of the antisymmetry
constraint. In random matrix 
theory, the limit $N \to \infty$
is taken at the end of the calculation.
The random matrix description is valid as long as energy
differences of the energy levels and wavefunctions of interest
are small compared to the Thouless energy 
$E_{\rm Th}$.\cite{kn:efetov1983,kn:altshuler1986}
For a disordered metal nanoparticle of size $L$, mean free path $l$, and 
Fermi velocity $v_F$, $E_{\rm Th} \sim v_F l/L^2$.

The parameter $\lambda$ in Eq.\ (\ref{eq:H0}) determines the
strength of the spin-orbit scattering, $\lambda^2 = \pi \gamma_{\rm so}/
\spacing = \pi/\tau_{\rm so} \spacing$, where $\tau_{\rm so} =
1/\gamma_{\rm so}$ is 
the spin-orbit
time and $\spacing$ is the mean spacing between (spin-degenerate)
eigenvalues of $H_{\rm GOE}$, see 
Refs.\cite{kn:brouwer2000,kn:matveev2000,kn:adam2002b}
The case $\lambda = 0$ corresponds 
to the absence of spin-orbit scattering, while the limit 
$\lambda \to \infty$ describes the situation when spin-rotation
symmetry is completely broken.
The factor $1/\sqrt{N}$ in front of $H_{\rm so}$ ensures that
the relation between $\lambda$ and the physical parameters 
$\tau_{\rm so}$ and $\spacing$ does not involve the matrix size $N$.

Each eigenvalue $\varepsilon_{\mu}$ of the Hamiltonian (\ref{eq:H0})
is doubly degenerate (Kramers degeneracy).
After diagonalization, the Hamiltonian $H_0$ can be written as
\begin{equation}
  H_0 = \sum_{\mu} \varepsilon_{\mu} 
  (\hat \psi^{\dagger}_{\mu1} 
  \hat \psi^{\vphantom{\dagger}}_{\mu1} +
  \hat \psi^{\dagger}_{\mu2} 
  \hat \psi^{\vphantom{\dagger}}_{\mu2}),
  \label{eq:H0single}
\end{equation}  
where $\hat \psi^{\dagger}_{\mu\alpha}$ and
$\hat \psi_{\mu\alpha}$
are creation and annihilation operators for an electron in the
state $|\mu \alpha\rangle$, where $\alpha=1,2$ labels the two
time-reversed states in the Kramers doublet.


Combining the single-electron Hamiltonian (\ref{eq:H0single}) and
the interaction Hamiltonian (\ref{eq:HintS}), one find the total
Hamiltonian
\be
  {\hat H} = \sum_{\mu} \varepsilon_{\mu} 
  (\hat \psi^{\dagger}_{\mu1} 
  \hat \psi^{\vphantom{\dagger}}_{\mu1} +
  \hat \psi^{\dagger}_{\mu2} 
  \hat \psi^{\vphantom{\dagger}}_{\mu2})
  - J {\bf {\hat S}}^2.
\label{hamiltonian}
\label{half_of_Hamiltonian}
\ee
Equation (\ref{hamiltonian}) is valid up to a charging energy that
depends on the electron number $N_e$ only; the charging energy plays
no role for in the problem we consider.

In the absence of spin-orbit interaction, the exchange interaction
(second term in Eq.\ (\ref{hamiltonian})) commutes with the
non-interacting Hamiltonian $H_0$ (first term in Eq.\
(\ref{hamiltonian})). The many-electron eigenstates are found by
diagonalizing $H_0$ at a fixed value of the total spin $S$ and its
$z$ component $S_z$. With spin-orbit interaction, however, the
interaction does not commute with $H_0$. 
This has important consequences for the
ground state and for the excitation spectrum of a metal nanoparticle.
Typically, for most normal-metal nanoparticles and for quantum dots,
$J$ is estimated to be in the range $0 \lesssim J/\spacing \lesssim
1$, see Sec.~\ref{sec:experiment}. 
While this implies that the effect of the exchange interaction
cannot be treated in first-order perturbation theory, 
interaction effects only cause correlations in a small window
around the Fermi energy which is, in principle, available to
direct numerical diagonalization. Hereto, we write the operator
$\vS$ for the total electron spin in terms of the creation and
annihilation operators $\hat \psi^{\dagger}_{\mu\alpha}$ and
$\hat \psi_{\mu\alpha}$ of the single-electron Hamiltonian $H_0$,
\begin{eqnarray}
  \hat \vS &=& \sum_{\mu,\nu} \sum_{\alpha,\beta=1,2}
  \hat \psi^{\dagger}_{\mu\alpha} \vs_{\mu\alpha,\nu\beta}
  \hat \psi_{\nu\beta}, \label{eq:Stotal}
\end{eqnarray}
where 
\begin{equation}
  (s_i)_{\mu\alpha,\nu\beta} = \frac{1}{2}
  \langle \nu\beta|\sigma_i|\mu\alpha \rangle,\ \ i=1,2,3.
\end{equation}

The quantity of interest in our calculation is the magnetic-field
dependence of the many-electron energy levels for small magnetic
fields and an odd number of electrons, which is described 
through the $g$ factors, see Eqs.\ (\ref{eq:V}) and (\ref{eq:Vg})
above. The 
magnetic-field dependence arises both through the Zeeman coupling
to the electron spin and through the orbital coupling to the angular
momentum.\cite{kn:matveev2000,kn:adam2002b} For large diffusive
metal nanoparticles and for not-too-large spin-orbit scattering strengths
$\tau_{\rm so} \spacing \gtrsim 1$,
the Zeeman coupling dominates.\cite{kn:matveev2000,kn:adam2002b}
Since the interaction effects studied here are most important for
$\tau_{\rm so} \spacing \sim 1$ (see below), we neglect the orbital
contribution to the $g$ factors in the discussion below. 
For a magnetic field $\vH$ along the $z$ axis, the Zeeman coupling to the 
magnetic field is described by the Hamiltonian
\begin{eqnarray}
  H_{\rm Z} &=& - 2 \mu_B H S_3,
\end{eqnarray}
where the $z$-component of the total spin $S_3$ is given by Eq.\
(\ref{eq:Stotal}) above.

\subsection{Tunneling spectroscopy}

Following Ref.\ \onlinecite{kn:vondelft2001},
we assume that the conductance of the tunneling contact connecting 
the nanoparticle to the source reservoirs is much smaller than the
conductance of the contacts connecting the particle to the drain
reservoir, so that the current $I$ through the nanoparticle is limited by 
the processes where an electron tunnels {\em onto} the particle.
In this case, one can assume that relaxation is sufficiently fast
that the nanoparticle is in the $N_e$-particle ground state 
$|N_E,0\rangle$
before each
tunneling event. It follows that current steps occur in the 
current-voltage
characteristic or, equivalently, a peak in the nanoparticle's differential
conductance $\partial I/\partial V$, when the source-drain
voltage $e V = e V_{k0} = E_{N_{\rm e} + 1,k} - E_{N_e,0}$, see
Eq.\ (\ref{eq:V}).
For a point contact that
injects electrons into the nanoparticle at position $\vr$, the size of the 
current step is proportional to the matrix element
\begin{eqnarray}
  w_{k} &=& | \langle N_e+1, k | \hat \psi^{\dagger}_{\uparrow}(\vr) 
  | N_e, 0 \rangle |^2 
  \nonumber \\ && \mbox{} + | \langle N_e+1, k | 
  \hat \psi^{\dagger}_{\downarrow}(\vr) | N_e, 0 \rangle |^2,
  \label{weights}
\end{eqnarray}
where the creation operator $\hat \psi^{\dagger}_{\sigma}(\vr)$
creates an electron with spin $\sigma$ in the nanoparticle at position $\vr$.
In terms of the basis of single-electron states, one has
\begin{equation}
  \hat \psi^{\dagger}_{\sigma}(\vr) =
  \sum_{\mu} \left(
  \hat \psi^{\dagger}_{\mu 1}
  \langle \vr \sigma | \mu 1 \rangle +
  \hat \psi^{\dagger}_{\mu 2}
  \langle \vr \sigma | \mu 2 \rangle \right).
  \label{creation_operator}
\end{equation}
In random matrix theory, the matrix element $\langle \vr \sigma | \mu
\alpha \rangle$, $\sigma=\pm 1$, is replaced by an (arbitrary) spinor in 
the $2N$-component vector representing the state $\mu \alpha$,
$\alpha=1,2$. 

\begin{figure}[t]
~\\
\epsfxsize= 0.98\hsize
\hspace{0.1\hsize}
\epsffile{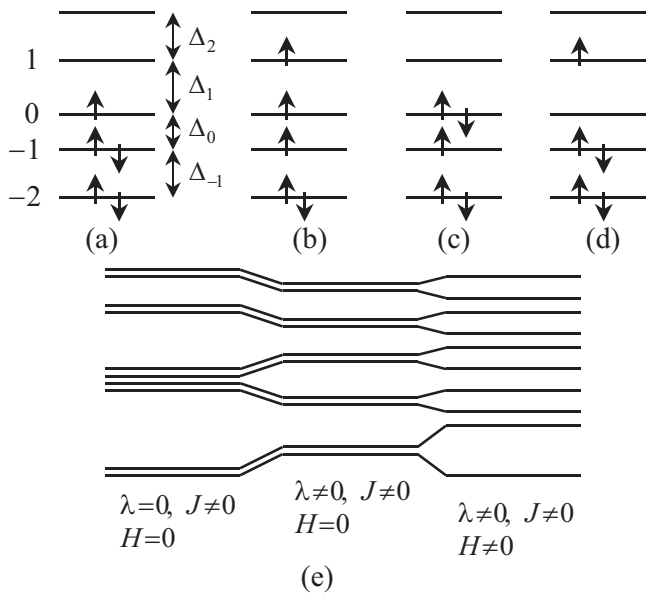}
\caption{ \label{fig:1}
Top panel: schematic representation of four lowest many-electron 
levels for a nanoparticle with an odd number of particles without
spin-orbit scattering. The spacings between the single-electron
levels are as indicated in the figure. The figure represents the case 
$0< \Delta_0 + \Delta_1 - 3J  < \Delta_0, \Delta_1$. Note that the
first excited state has spin $S=3/2$ and is fourfold degenerate.
Bottom panel: Evolution of energy levels of the nanoparticle for small 
spin-orbit scattering rates $\lambda$ and magnetic fields $H$. 
Spin-orbit scattering separates the quadruplet (b) into two
doublets with ill-defined spin. The magnetic field lifts the
degeneracy of all remaining doublets.}
\end{figure}

In the presence of both spin-orbit scattering and electron-electron
interactions, the $N_e$-electron states $|N_e,l\rangle$
are non-degenerate for zero magnetic field if $N_e$ is even. 
In that case, $\partial E_l/\partial H = 0$ at $H=0$.
On the other hand, $(N_e+1)$-electron states are doubly degenerate 
if $N_e$ is even. These states split in a magnetic field. For small 
magnetic fields the splitting is linear. Hence, if $N_e$ is even,
the $g$ factor $g_{kl}$ associated with voltage $V_{kl}$ at which 
a current steps occurs is directly related to the magnetic-field 
derivative of the $(N_e+1)$-electron level $|N_e + 1,l\rangle$,
\begin{equation}
  \frac{\partial E_{N_{\rm e} + 1,k}}{\partial B}
  = \pm \frac{1}{2} g_{k0} \mu_B,\ \
  \mbox{$N_{\rm e}$ even}.
\end{equation}
In the remainder of this paper we continue to refer to the
even-electron ground state as the ``$N_e$-electron ground state''
and to the odd-electron states as the ``$(N_e+1)$-electron states''.

An example showing how the various degeneracies are lifted by
the exchange interaction and by a magnetic field is shown in Fig.\
\ref{fig:1}. The top panel of the 
figure shows the four lowest many-electron states
for an odd number of electrons, for the specific case that the
ground state has spin $S=1/2$ and the lowest excited state has
$S=3/2$ in the absence of spin-orbit scattering. Without spin-orbit
scattering, the $S=3/2$ state is fourfold degenerate. Spin-orbit
scattering lifts the fourfold degeneracy of the $S=3/2$ quadruplet,
separating it into two doublets with ill-defined spin (lower panel
of Fig.\ \ref{fig:1}, center). Finally, an applied magnetic field
lifts the degeneracy of all doublets. For an even electron number,
spin-orbit scattering lifts all possible degeneracies; to first
order in the field, the applied magnetic field has no effect.

The definition that $g$ factors are derivatives of energy levels to
the magnetic field implies that the Zeeman energy scale is the
smallest nonzero energy scale in the problem.  In particular, it is
smaller than the spin-orbit induced splittings of high-spin states
(see Fig.\ \ref{fig:1}), these splittings being proportional to
the spin-orbit scattering rate $\lambda^2 \propto
1/\tau_{\rm so}$. However, when
$g$ factors are calculated without spin-orbit scattering, the Zeeman
energy is (by definition) 
larger than the spin-orbit rate. We'll find below that the
two limits do not commute in the presence of electron-electron
interactions, {\em i.e.,} that $g$ factor distributions calculated in
the limit $\lambda \to 0+$ 
are different from the $g$ factors at
$\lambda=0$. (Without spin-orbit scattering all peaks in the
differential conductance split with $g$ factor $g=2$.)  It should be
pointed out that experiment involves finite magnetic fields, for which
the Zeeman energy can be larger than the spin orbit rate. For such
magnetic-field dependences of the levels, a different slope at zero
field can easily go unnoticed.

Although $g$ factors contain information on the {\em positions} 
of peaks in the tunneling spectrum of the nanoparticle, knowledge of the 
{\em sizes} of the
peaks in the differential conductance is important for a correct
interpretation of the results. Peak heights contain information that
would be formulated in terms of ``selection rules'' in the absence
of spin-orbit scattering. Indeed, without spin-orbit scattering, the
total spin $S$ and its $z$ component $S_3$ can change by $1/2$ at
most in a tunneling process, which limits the possible transitions
between $N_e$ and $N_e+1$ electron states, and, hence, the possible
locations of peaks in the differential conductance. It is because of
the selection rules that one does not observe $g$ factors larger than
two, despite the fact that there exist high-spin many-electron states.
Similarly, without interactions, the occupation of single-electron
levels cannot change by more than one electron, which also limits the
number of allowed transitions in tunneling spectroscopy.
With spin-orbit
scattering and interactions, peaks that were previously ``forbidden'' 
are present, in principle, although their height may be small. We
return to this issue in more detail in Sec.\ \ref{sec:simulations}.

\begin{figure}[t]
\epsfxsize= 0.9\hsize
\hspace{0.1\hsize}
\epsffile{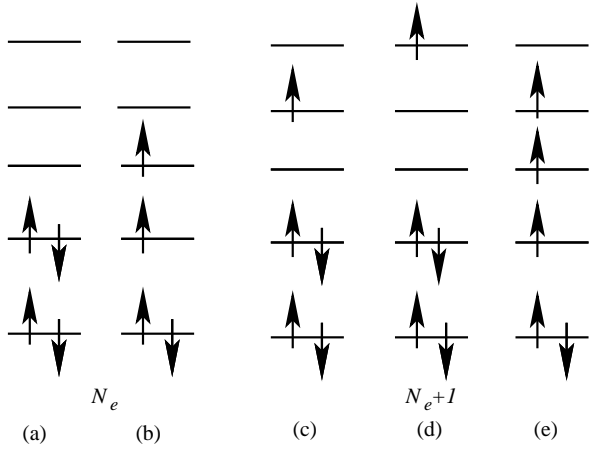}
\caption{ \label{fig:2}
Left: Schematic representation of ground states for an even number of 
electrons without spin-orbit scattering. Depending on the strength
of the exchange interaction, the $N_e$-electron ground state may
have spin $S=0$ (a) or $S=1$ (b). Right: three excited
$(N_e+1)$-electron states.}
\end{figure}

The example in Fig.\ \ref{fig:2} may further clarify the role of 
selection rules. The figure shows two possible $N_e$-electron
ground states (left) and three $(N_e+1)$-particle excited states
(right).
Without spin-orbit scattering or without exchange interactions,
the states (c) and (d) can be accessed from the $N_e$-particle
ground state (a), but not state (e). Similarly, state (e) can
be accessed from ground state (b), but not (c) and (d). When
spin-orbit scattering and exchange interactions are both present,
the $N_e$-electron ground state is a superposition of the states
(a) and (b) and all possible transitions have a finite matrix 
element. Notice that, 
since the energy difference between states (a) and (b)
is typically small --- on average, equal to $\delta - 2 J$, 
mixing of these two states is strong for spin-orbit scattering
rates $\lambda \sim 1$.

\subsection{Matrix elements of interaction Hamiltonian}

For the construction of the interaction Hamiltonian in the basis of 
many-electron eigenstates of the Hamiltonian $H_0$, one needs 
explicit equations for the matrix elements of the exchange
interaction in that basis. Since the exchange interaction changes
the single-electron states of at most two electrons,
the only nonzero matrix elements of the exchange interaction
occur between states that can be written in the
form 
$\hat \psi^{\dag}_{\mu\alpha} \hat \psi^{\dag}_{\nu\beta}|F\rangle$
and 
$\hat \psi^{\dag}_{\mu'\alpha'} \hat
\psi^{\dag}_{\nu'\beta'}|F\rangle$,
where $|F\rangle$ a certain reference non-interacting state. After
some algebra one then finds
\begin{widetext}
\begin{eqnarray}
  \lefteqn{
  \langle F|\hat \psi_{\nu'\beta'} \hat
  \psi_{\mu'\alpha'}  \vS^2 
  \hat \psi^{\dag}_{\mu\alpha} \hat \psi^{\dag}_{\nu\beta}|F\rangle }
  \nonumber \\
  &=& \vphantom{\sum_{\gamma \in F}}
  \left
  (\delta_{{\mu\alpha},{\mu'\alpha'}}\delta_{{\nu\beta},{\nu'\beta'}}
  - 
  \delta_{{\nu\beta},{\mu'\alpha'}}\delta_{{\mu\alpha},{\nu'\beta'}}\right )
  \langle F| \vS^2 | F \rangle 
  - 2 \left [
{\bf s}_{{\nu'\beta'}{\mu\alpha}}\cdot{\bf s}_{{\mu'\alpha},{\nu\beta}} -
{\bf s}_{{\mu'\alpha'},{\mu\alpha}}\cdot{\bf s}_{{\nu'\beta'},{\nu\beta}}
  \right ]  
  \nonumber\\ && \mbox{} 
  - 2 \delta_{{\nu\beta},{\nu'\beta'}} 
\sum_{{\phi,\gamma}\in { F}}
\left [ 
{\bf s}_{{\mu'\alpha'},{\phi\gamma}}\cdot{\bf s}_{{\phi\gamma},{\mu\alpha}} -
{\bf s}_{{\mu'\alpha'},{\mu\alpha}}\cdot{\bf s}_{{\phi\gamma},{\phi\gamma}}
  \right ] 
  - 2\delta_{{\mu\alpha},{\nu'\beta'}} 
\sum_{{\phi,\gamma}\in { F}}
\left [ 
{\bf s}_{{\mu'\alpha'},{\nu\beta}}\cdot{\bf s}_{{\phi\gamma},{\phi\gamma}} -
{\bf s}_{{\phi\gamma},{\nu\beta}}\cdot{\bf s}_{{\mu'\alpha'},{\phi\gamma}}
 \right ]  
  \nonumber\\ && \mbox{} - 
  2 \delta_{{\nu\beta},{\mu'\alpha'}} 
\sum_{{\phi,\gamma}\in { F}}
\left [ 
{\bf s}_{{\nu'\beta'},{\mu\alpha}}\cdot{\bf s}_{{\phi\gamma},{\phi\gamma}} -
{\bf s}_{{\phi\gamma},{\mu\alpha}}\cdot{\bf s}_{{\nu'\beta'},{\phi\gamma}}
  \right ]
  - 2 \delta_{{\mu\alpha},{\mu'\alpha'}} 
\sum_{{\phi,\gamma}\in { F}}
\left [ 
{\bf s}_{{\phi\gamma},{\nu\beta}}\cdot{\bf s}_{{\nu'\beta'},{\phi\gamma}} -
{\bf s}_{{\nu'\beta'},{\nu\beta}}\cdot{\bf s}_{{\phi\gamma},{\phi\gamma}}
\right ],
\nonumber \\
\label{matrix_element}
\end{eqnarray}
where
\be
\langle F | \vS^2 | F \rangle =
  \frac{3}{4}(N_e-2) -
  2 \sum_{{\phi,\gamma} \in F}
  \sum_{{\phi',\gamma'} \in F}
  \left [
  \vs_{\phi \gamma, \phi'\gamma'}\cdot{\bf s}_{\phi'\gamma',\phi \gamma }
   - 
  \vs_{\phi \gamma, \phi \gamma }\cdot{\bf s}_{\phi'\gamma',\phi'\gamma'} 
  \right ].
\label{diagonal_matrix_element}
\ee
\end{widetext}%
The summations over $\phi$ and $\gamma$ extend over all
single-electron states $|\phi,\gamma\rangle$, $\gamma=1,2$, that are
occupied in the state $|F\rangle$.

Similarly, matrix elements of the Zeeman energy $H_{\rm Z}$
are nonzero only if many-electron states differ by not more than
one electron, {\em i.e.,} between states of the form
$\hat \psi^{\dag}_{\mu\alpha}|F\rangle$ and 
$\hat \psi^{\dag}_{\nu\beta}|F\rangle$. For those states, one
needs the matrix elements
\be
  \langle F| \hat \psi_{\nu\beta} \vS
  \hat \psi^{\dag}_{\mu\alpha} |F\rangle
  =
  \vs_{\nu\beta,\mu\alpha} +
  \delta_{\nu\beta,\mu\alpha} \sum_{ \phi,\gamma \in F}
  \vs_{\phi\gamma,\phi\gamma}.
\label{magnetic_field_matrix_element}
\ee

At this point, it is important to verify the applicability of the
random matrix theory. In order for random matrix to apply, summations
over the Fermi sea should converge within a Thouless energy from the
Fermi level. In Eqs.\ (\ref{matrix_element}),
(\ref{diagonal_matrix_element}) and
(\ref{magnetic_field_matrix_element}), the sums of the form
$\sum_{\phi,\gamma \in F} \vs_{\phi\gamma,\phi\gamma}$ clearly satisfy
this condition, by virtue of the equality 
\be
  \vs_{\phi 1,\phi 1} + \vs_{\phi 2, \phi 2}
  = 0,
\label{identity}
\ee
which follows from the observation that the states $|\phi,1\rangle$ 
and $|\phi,2\rangle$ are time reversed. The sums
of the form $\sum_{\phi,\gamma \in F} \vs_{\mu\alpha,\phi\gamma} \cdot
\vs_{\phi\gamma,\nu\beta}$ also meet this condition, since the summand
decreases $\propto 1/(\varepsilon_F - \varepsilon_{\phi})^2$ for
$\phi$ well below the Fermi level and $\mu$ and $\nu$ close to the
Fermi level.\cite{kn:adam2002b} In the diagonal matrix element
(\ref{diagonal_matrix_element}) the sum over $(\phi,\gamma)$ and
$(\phi',\gamma')$ is logarithmically divergent as a function of the
Fermi energy. This, however, has no consequences for the
magnetic-field dependence of the many-electron states and the peak
heights, since the divergence is the for all matrix elements and
simply corresponds to the overall shift of the ground state energy. We
conclude that random matrix theory can be used to access the
many-electron ground state and the low-lying excited states.

\section{Results and discussion}
\label{sec:simulations}

In order to calculate $g$ factors and peak heights, we have 
diagonalized the Hamiltonian (\ref{hamiltonian}) numerically.

Our numerical procedure is as follows:
We first diagonalized the non-interacting Hamiltonian $H_0$. The
interaction is considered in a truncated basis of the many-electron
states, taking the 92 lowest lying many-electron eigenstates of
$H_0$ for a $N_e+1$ electrons ($N_e$ even), and the 76 lowest lying
states for $N_e$ electrons. Finally, we diagonalized
the truncated interaction and found the $g$ factors of the $M=8$
lowest-lying $(N_e+1)$-electron states, and the peak heights that
follow for transitions from the $N_e$-electron ground state. 
For the calculation of the $g$ factors we introduce a small 
magnetic field and calculate the magnetic field derivative
numerically. We
verified that truncating the interaction Hamiltonian at the 
lowest lying 92 and 76 many-electron states has no effect on the
final results by comparing our results to those that were obtained
using a smaller basis set.

We have investigated exchange-interaction strength ranging from
$J=0$ to $J = 0.6\spacing$ which is valid for most metals,
see Sec.~\ref{sec:experiment}.
The same parameter range should apply to quantum dots.
Analysis of the Coulomb blockade 
peak spacing distribution\cite{kn:usaj2003} suggests that
$J/\spacing$ is between $0.3$ and $0.4$
in large quantum dots in a GaAs/GaAlAs
heterostructure; whereas recent density functional
studies of ground state spin distributions in ballistic quantum
dots are compatible with the Hamiltonian (\ref{hamiltonian})
only if $J\approx 0.6\spacing$, see Ref.\ \onlinecite{kn:jiang2003}.

Important changes occur within the range of exchange interactions we
address here. Since the metal nanoparticle is assumed to relax to the
$N_e$-particle ground state between tunneling events, the
(statistical) properties of the $N_e$-particle ground state play a key
role in determining the $g$ factor distribution. Without spin-orbit
scattering for $J \lesssim 0.3\spacing$, there is only a small
probability that the $N_e$-particle ground state has spin one, and a
vanishing probability that the $N_e+1$-particle ground state has spin
$3/2$ ($N_e$ is assumed even).\cite{kn:brouwer1999b,kn:oreg2002} The
probability to find an $N_e$-particle ground state with $S=1$ becomes
appreciable for $J \gtrsim 0.3$, whereas the probability to find an
$(N_e+1)$-particle ground with spin $3/2$ becomes significant for $J
\gtrsim 0.5\spacing$ only. For the values of $J$ we consider, states
with spin $\ge 5/2$ do not play a role; they have been excluded
from the truncated many-electron basis.

The strength of spin-orbit scattering strength $\lambda$ is taken 
from $0$ to $2.8$. Although larger spin-orbit scattering strengths
do occur in metal nanoparticles,\cite{kn:petta2001,kn:petta2002} interaction
effects are small at those values of $\lambda$ and the non-interacting
theory of Refs.\
\onlinecite{kn:brouwer2000,kn:matveev2000,kn:adam2002b} works well.

The random matrices in our simulation are taken of size $2N=400$.
This ensures that the condition ${\lambda}^2 \ll N$ necessary for
the applicability of the random matrix (\ref{eq:H0}) is satisfied
for all values of $\lambda$. For $\lambda <2 $ we have taken $2N=200$
in the simulations. 

\subsection{Average $g$ factors}
\label{sec:threshold}

\begin{figure}
\epsfxsize= 0.75\hsize
\epsffile{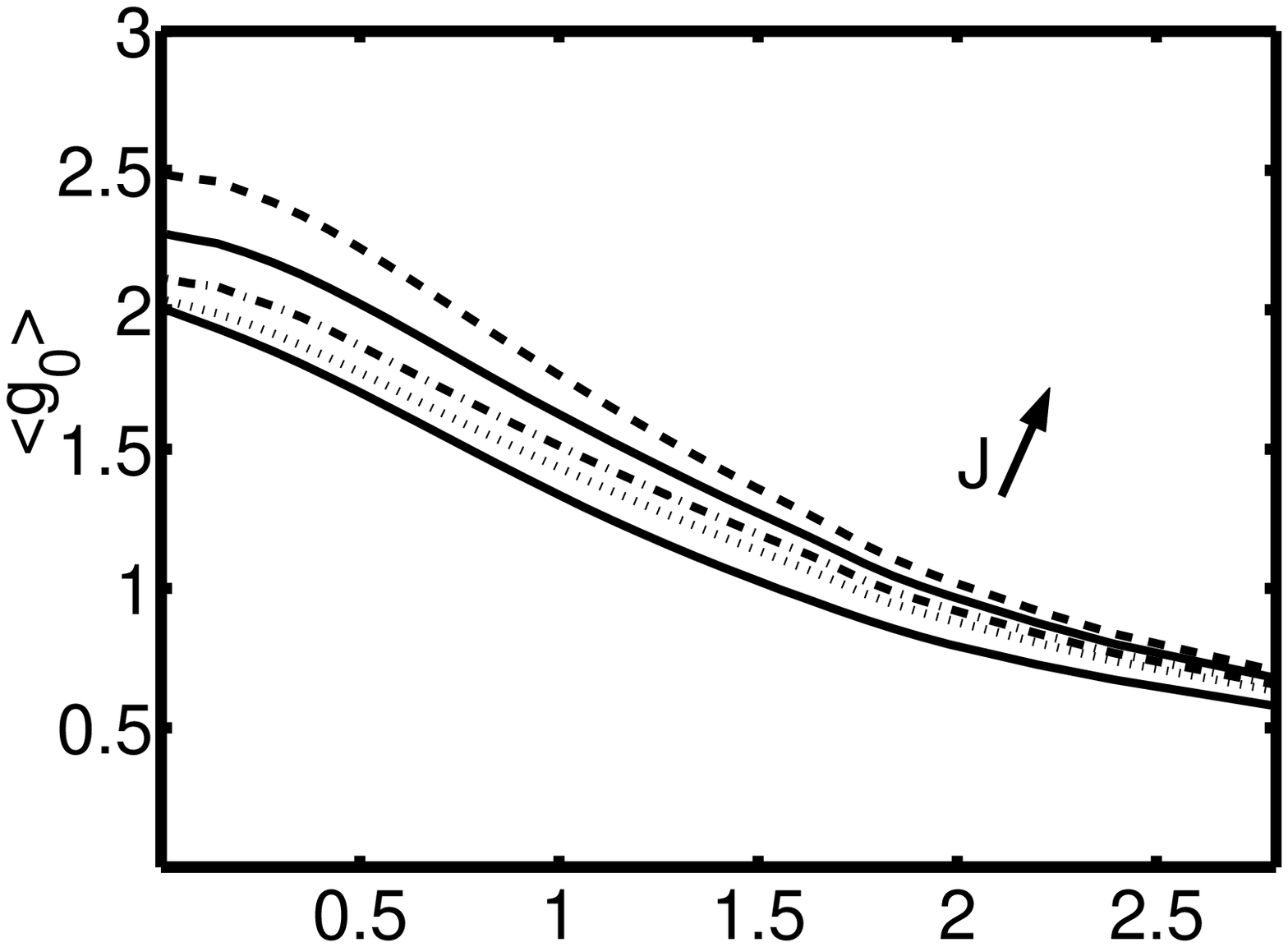}

\epsfxsize= 0.75\hsize
\epsffile{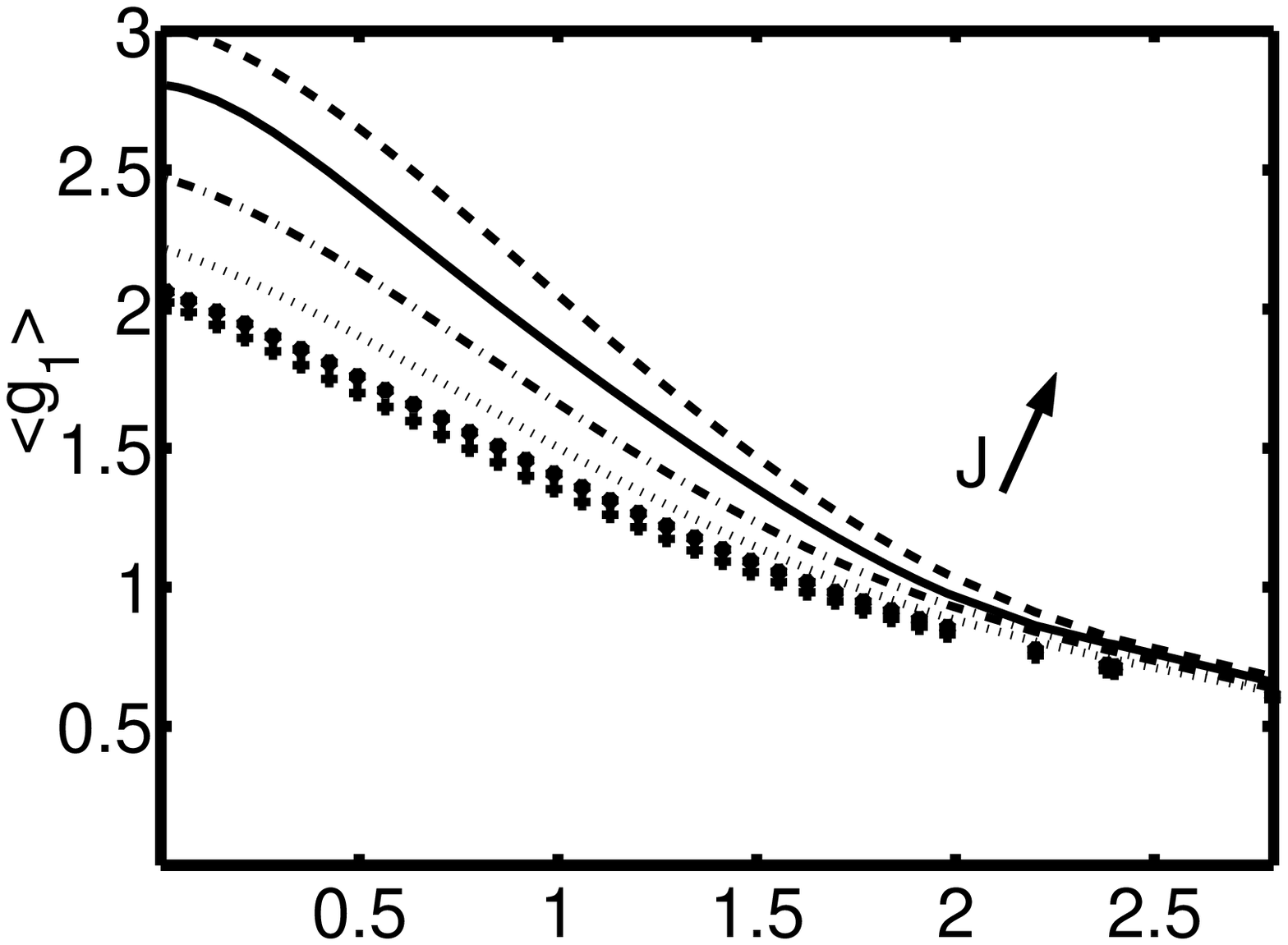}

\epsfxsize= 0.75\hsize
\epsffile{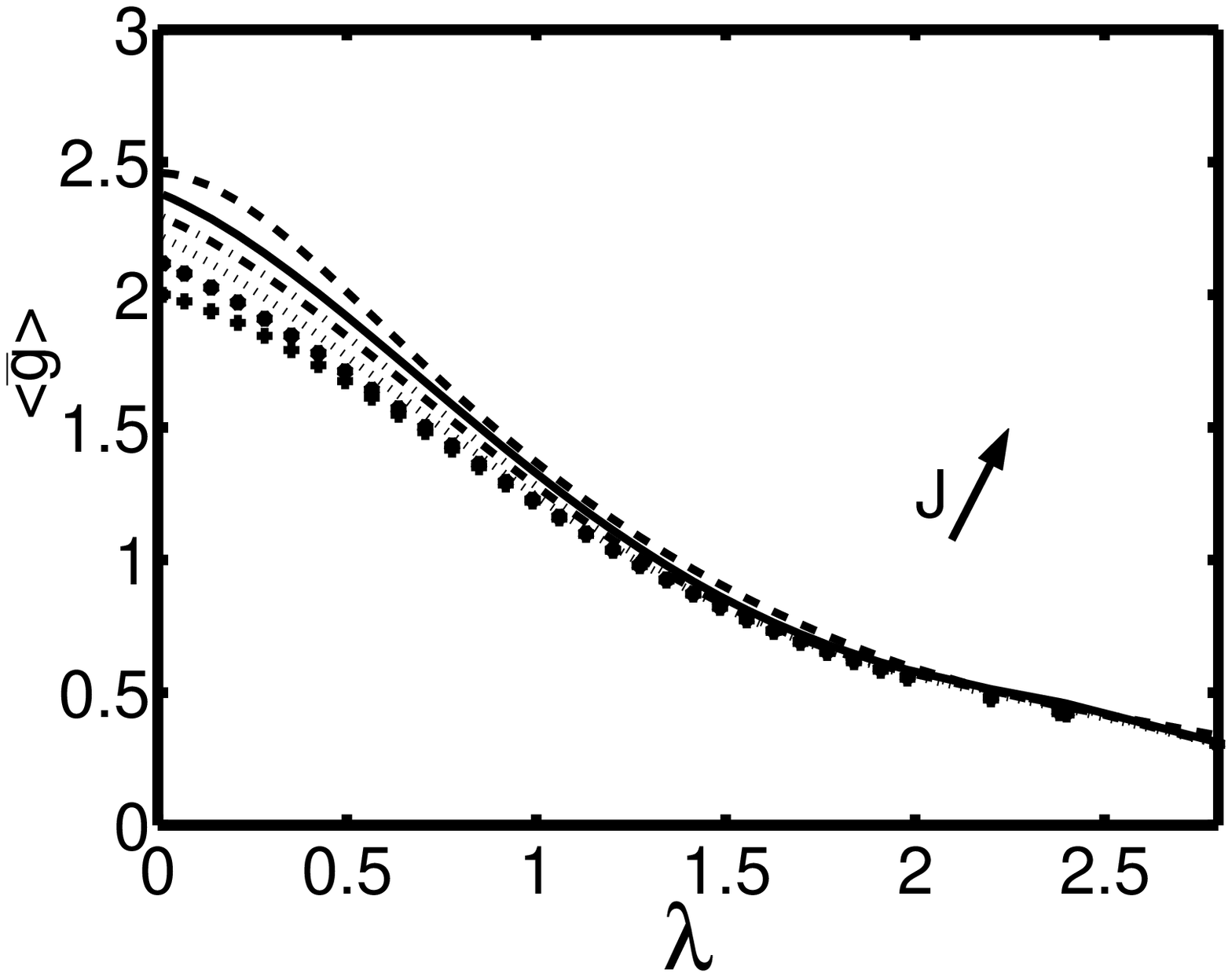}
\caption{Ensemble averaged $g$-factors.
 Top: average $g$ factor
$\langle g_0 \rangle$ of $(N_e+1)$-particle ground state for
$J = 0$ (the lowest solid curve), $J = 0.3\spacing$, $J = 0.4\spacing$,
$J = 0.5\spacing$, and $J = 0.6\spacing$
Middle:
average $g$ factor $\langle g_1 \rangle$ of first 
excited $(N_e+1)$-particle state for $J/\delta = 0.1, 0.2, 0.3, 0.4, 0.5$
and 0.6.
Bottom: ensemble averaged $g$
factor averaged over the first $M=8$ $(N_e+1)$-electron states
for  $J/\delta = 0.1, 0.2, 0.3, 0.4, 0.5$ and 0.6.
} 
\label{ground_state_g_factor}
\end{figure}

We have calculated ensemble averages of the $g$ factors $\langle
g_k\rangle$,
$k=0,1,\ldots,M-1$ of the $M$ lowest $(N_e+1)$-electron states.
Here $g_k$ is the $g$ factor corresponding to the $k$th
$(N_e+1)$-electron system, $k=0,1,\ldots,M-1$. The ensemble average
is taken over 300 realizations.
In Fig.~\ref{ground_state_g_factor} we show the ensemble averaged 
$g$ factors
for the ground state and the first excited state, $\langle g_0 \rangle$
and $\langle g_1 \rangle$, as well as the average over all
calculated $g$ factors $\langle \bar g \rangle$
\begin{equation}
  \bar g 
 = M^{-1} \sum_{k=0}^{M-1}  g_k .
\end{equation}

The first observation to be made from Fig.\
\ref{ground_state_g_factor} is that for $J \gtrsim 0.4\spacing$ and 
$\lambda \lesssim 2$ interactions 
lead to a significant increase in the average $g$ factor. In fact,
there is a significant parameter range for which $\langle \bar g 
\rangle > 2$. The origin of the large $g$ factors is that
exchange interactions lift the degeneracy with respect to the 
total spin $S$. Hence, with exchange interactions, many-electron
states with a finite spin are energetically separated from states
with spin $0$ or $1/2$. For the parameter range considered here,
the relevant nontrivial spin states have $S=3/2$ for an odd number
of electrons. The role of spin-orbit scattering is to lift the
fourfold degeneracy of the $S=3/2$ states and, for larger spin-orbit
strengths, to suppress the spin content of the single-electron states
that build the many-electron state. Let us first discuss the effect
of lifting the degeneracy of the $S=3/2$ state by spin orbit
scattering.

In general, spin orbit scattering splits the fourfold degenerate 
$S=3/2$ state into two doublets. Neglecting contributions from 
other many-electron states, each doublet consist of two states 
that can written in the form 
\be
  |\mbox{state}\rangle = \sum_{n=-3/2}^{3/2} a_{n}
  |3/2,n\rangle, \label{eq:doublet}
\ee      
and the time-reversed of Eq.\ (\ref{eq:doublet}), which is
obtained by sending $a_n \to \mbox{sign}(n) a_{-n}^*$.
Here $|S,S_3\rangle$ is the $(N_e+1)$-electron state with 
total spin $S$ and $z$ component
of the spin equal to $S_3$. Because the spin-orbit matrix elements
are random, the amplitudes $a_n$ are essentially random as well.
(This statement is verified in the appendix.)
The $g$ factor of the state (\ref{eq:doublet}) 
is
\begin{eqnarray}
  g^2 = \left( \sum_{n=-3/2}^{3/2} 4 n |a_n|^2 
  \right)^2 + \left| \sum_{n=-3/2}^{3/2} 4 |n| a_n a_{-n}
  \right|^2.
\end{eqnarray}
One easily verifies that this can be larger than two. 
With exchange interaction but without spin-orbit scattering, 
there is a finite probability
that the $N_e$-particle ground state ($N_e$ even) has spin $S=1$.
In that case, it has two singly occupied orbitals and, hence, in
principle, a finite overlap with a state of the form
(\ref{eq:doublet}) after addition of an electron. With spin-orbit
scattering, the $N_e$-electron ground state is guaranteed to be
non-degenerate, so that its derivative to the magnetic field is 
zero. We conclude that, the $g$ factor of the state (\ref{eq:doublet}) 
can be larger that two, that it can correspond to a transition between
the $N_e$-electron ground state and an $(N_e+1)$-electron state,
and that the corresponding
conductance peak has a finite height.

A finite amount of spin-orbit scattering stabilizes the above arguments
by increasing the splittings between many-electron states that are
degenerate in the absence of spin-orbit coupling. On the other hand,
with moderate
spin-orbit scattering, more many-electron states are added in 
the doublet (\ref{eq:doublet}). This has two consequences: (1) the the
spin content of each of the underlying single-electron states is
reduced, which, eventually, leads to a suppression of $g$ factors, and
(2) when more many-electron states are admixed, overlaps and, hence,
peak heights are increased, so that the role of selection rules is
further diminished. In order to illustrate the value of $\lambda$
needed to admix different many-electron states, we note that for the
$N_e$-particle ground state without spin-orbit interaction, the energy
separation between the $S=1$ and $S=0$ ground states is $2J-\delta$ on
average. Hence, even for a relatively small spin-orbit scattering rate
$\lambda \sim 0.5$ the ground state with spin-orbit scattering will
have significant weight in both of these states.

Also notice that, unlike in the non-interacting case, the average $g$ 
factor
depends on the excitation energy of the many-electron state for 
$\lambda \lesssim 2$, see Fig.\ \ref{ground_state_g_factor}.
The origin of this dependence is that, without
spin-orbit scattering, the probability of that an
$(N_e+1)$-particle state has nontrivial spin ($S \ge 3/2$) increases
with the excitation energy. As discussed above, although 
spin-orbit scattering lifts the fourfold degeneracy of these states
and suppresses the spin, it is the underlying nontrivial spin
character persisting to finite $\lambda$
that gives rise to the increased $g$ factors.

A remarkable feature of Fig.\ \ref{ground_state_g_factor} is that
the ensemble averaged $g$ factor $\langle \bar g \rangle$
does not approach two in the limit
$\lambda \to 0$. On the other hand, without spin-orbit scattering,
all observed $g$ factors should equal $g=2$. In Sec.\ 
\ref{sec:theoretical_description}
we discussed why there can be a difference between $g$ factors in
the limit $\lambda \to 0$ and $g$ factors calculated without
spin orbit scattering, {\em i.e.,} at $\lambda=0$. In fact, in Fig.\
\ref{ground_state_g_factor}, $\langle \bar g \rangle$ is 
overestimated for $\lambda \to 0$, because the plain ensemble average 
does not take tunneling spectroscopy peak heights or ``selection 
rules'' into
account: The average is taken over all $(N_e+1)$-particle states
irrespective of the height of the corresponding peak in the 
differential conductance. In particular for small $J$, one 
expects that $(N_e+1)$-electron states with $g$ factors larger
than two are likely to
have small tunneling spectroscopy peak heights.

\begin{figure}
\epsfxsize= 0.75\hsize
\epsffile{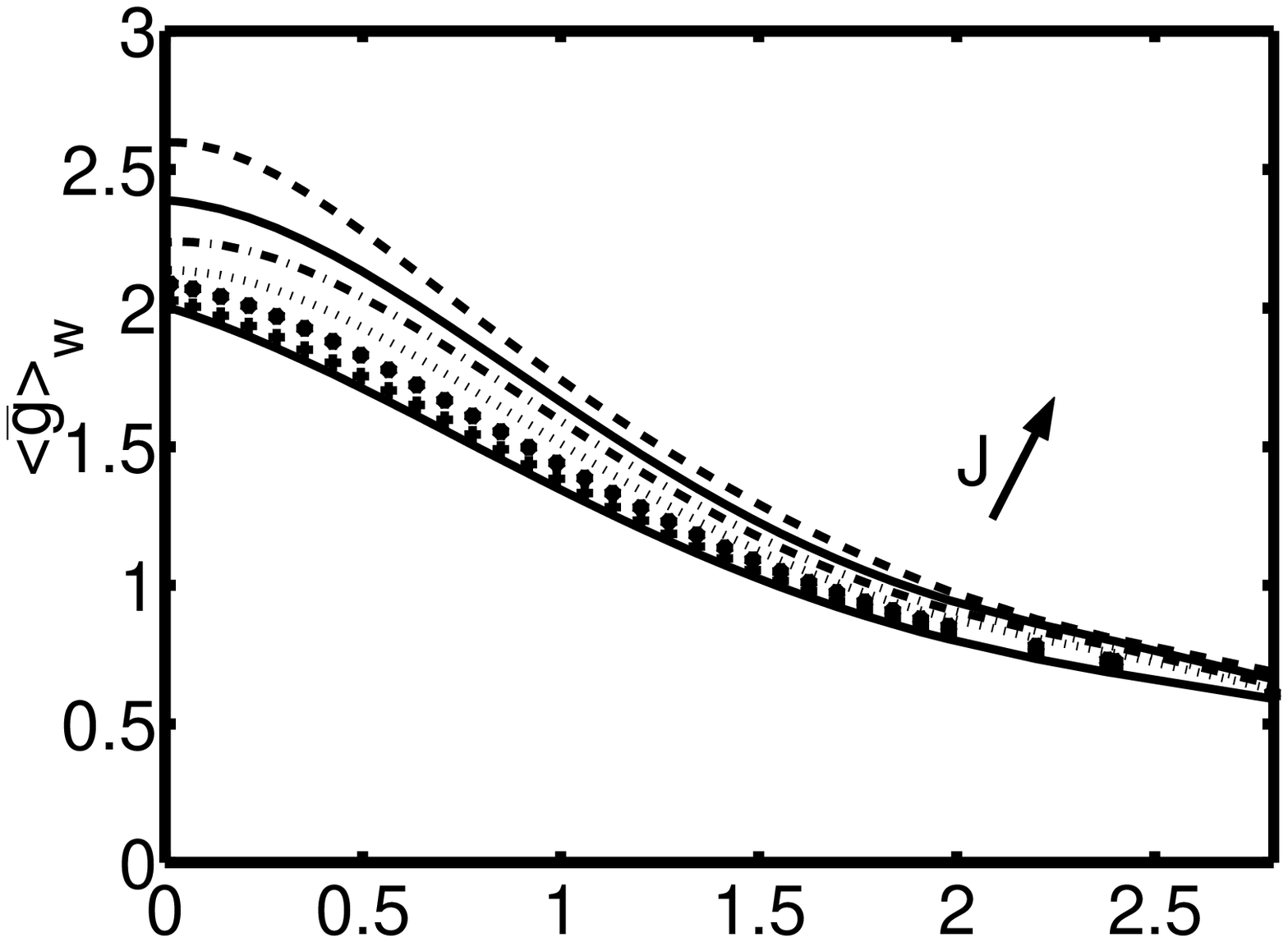}
\epsfxsize= 0.75\hsize
\epsffile{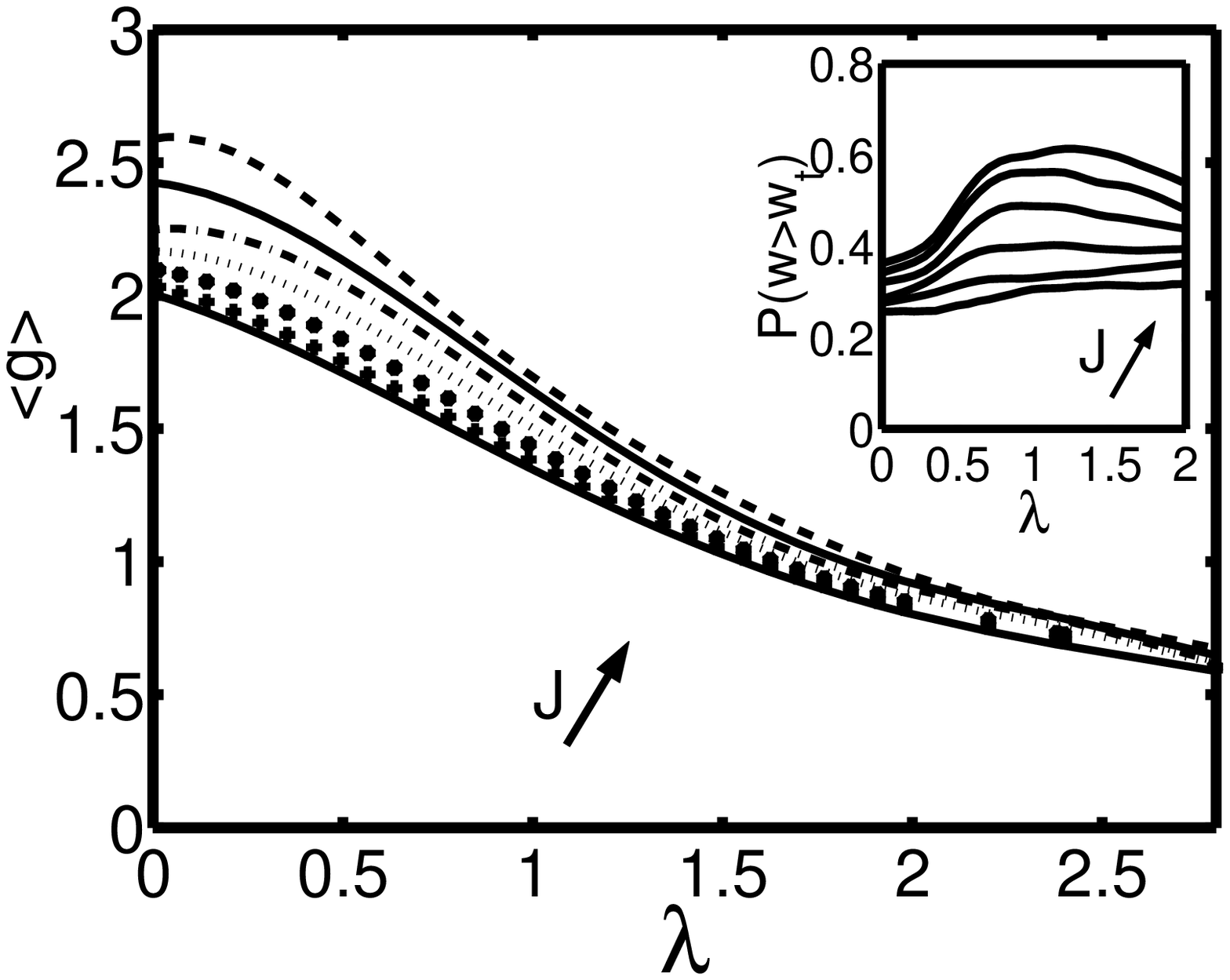}
\caption{Top: average of all calculated $g$ factors, where each 
$g$ factor is weighed by its normalized peak height
  (\protect\ref{eq:wg}). Bottom: average of all calculated $g$
factors for which the normalized peak height is larger than $0.1$.
In both panels, results are shown for $J/\spacing=0$, $0.1$, $0.2$,
$0.3$, $0.4$, $0.5$, and $0.6$.
For $J=0$ there is no difference with Fig.\
\protect\ref{ground_state_g_factor}.
Inset of lower panel: the probability for a level
to be visible in the experiment, 
i.e. to have a weight larger than the threshold
one, see 
(\protect\ref{weight_criterion}).} 
\label{weighted_average_g_factors}
\end{figure}

We have taken two different approaches in order to account for 
``selection rules''. First, we have replaced
the ensemble average by a ``weighed'' average, in which every 
$g$ factor is weighed by the normalized peak height
\begin{eqnarray}
  \langle \bar g \rangle_{\rm w} &=&
  \left\langle \sum_{k=0}^{M-1} \tilde w_k g_k \right\rangle,
  \ \
  \tilde w_k = \frac{M w_k}{\sum_{k=0}^{M-1} w_k}.
  \label{eq:wg}
\end{eqnarray}
In the second approach, we have removed all peaks with normalized
weight $\tilde w_k$ below a certain threshold value,
where we arbitrarily set the threshold to $\tilde w_{\rm t} =
0.1\times \max^{M}_{k=1} w_k$,
\begin{equation}
  \tilde w_k \to \tilde w_{k,{\rm t}} =
  \left\{ \begin{array}{ll} 1 & \mbox{if $ w_k
  \ge w_{\rm t}$}, \\
  0 & \mbox{if $w_k < w_{\rm t}$}. \end{array} \right.
\label{weight_criterion}
\end{equation}
In this method, the number of levels per realization depends on the 
realization,
\begin{equation}
  M_{\rm t} = \sum_{k} \tilde w_{k,{\rm t}}
\end{equation}
and the average $g$-factor is determined through
\begin{equation}
 \langle \bar g \rangle = \left \langle \frac{1}{M_t}
\sum_{k=0}^{M-1} \tilde w_{k,t} g_k\right \rangle .
\end{equation}
The threshold mimics the experimental reality that small peaks cannot
be distinguished from the noise, and, hence, have their $g$ factors
left out in the statistical analysis. While the second method is
closer to the way experiments are analyzed --- all $g$ factors of
conductance peaks that are observed are taken equally into account
in the average --- the first method has the advantage that it does
not contain the somewhat arbitrary threshold at $\tilde w_k = \tilde
w_{\rm t} = 0.1$. 
Both methods enforce the ``selection rules'' in the absence of
spin-orbit scattering. They also give almost identical results for
the average $g$ factor, as is seen from Fig.\
\ref{weighted_average_g_factors} where we show the weighted
ensemble average of the $g$ factors of all levels considered 
$\langle \bar g \rangle$, as well as the ensemble average
calculated using the ``threshold'' method.
As shown by comparison of Figs.\ \ref{weighted_average_g_factors}
and \ref{ground_state_g_factor}, in the limit $\lambda \to 0$,
the average $g$ factors are close to two for small $J$, whereas
$\langle \bar g \rangle$ is significantly higher than two for
$J \gtrsim 0.3$. In the inset of the lower panel of 
Figs.\ \ref{weighted_average_g_factors} we show the probability
of the level to be visible, i.e. to have a peak height above
the threshold. Remarkably, the curves for $J\agt 0.3\delta$ 
have a maximum for moderate values of spin-orbit
scattering $\lambda\simeq 1$. This has a direct physical interpretation:
both for small and large $\lambda$, 
approximate selection rules are in place. For small $\lambda$
these selection rules represent the conservation of spin 
at $\lambda = 0$, whereas for large $\lambda$ they follow from
the suppression of the (exchange) interaction, which causes the remaining
physics to be single-particle like.

\subsection{fluctuations of $g$ factors}

\begin{figure}
\epsfxsize= 0.75\hsize
\epsffile{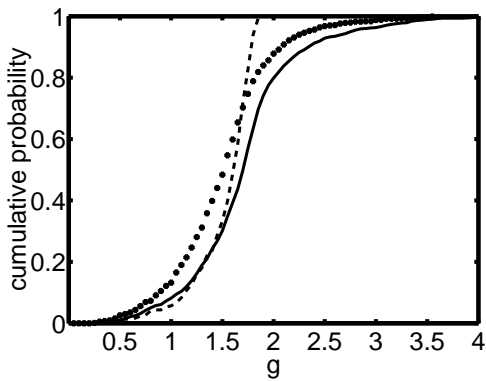}
\caption{Cumulative probability distributions function for $g$-factors
for the cases $\lambda = 0.7$, $J=0$ (dashed curve)
$\lambda = 0.7$, $J=0.3$ (solid), and $\lambda=0.9$, $J=0.3$
(dotted). Even a relatively
small strength of the exchange interaction is enough in order
to broaden the distribution function significantly.
} 
\label{probability_distribution}
\end{figure}
Cumulative probability distributions of $g$ factors are shown in
Fig.\ \ref{probability_distribution} 
for $J=0$, $\lambda=0.7$, for $J=0.3\spacing$,
$\lambda=0.7$ and for $J=0.3\spacing$, $\lambda=0.9$
(we have taken into account only peaks whose weights are nonzero
according to the criterion (\ref{weight_criterion})). 
 Comparing
the two distributions at $\lambda=0.7$, one notices that the
exchange interaction has little effect on the tail of the $g$ factor
distribution for very small $g$ factors. However, for larger
$g$ factors, the weight of the probability distribution is shifted
towards larger $g$ factors, including a long tail in the region
$g > 2$. Fig.\ \ref{probability_distribution} confirms the 
previous observation that the effect of exchange interactions is to 
increase the average $g$ factors. The spin-orbit scattering rate
for the third probability distribution shown in Fig.\ 
\ref{probability_distribution}, 
$\lambda=0.9$, has been chosen such that the average $g$ factor
$\langle {\bar g} \rangle_t \approx 1.58$
 coincides
with that of the case $J=0$, $\lambda=0.7$. Comparing the two
probability distributions, we conclude that the interactions 
still
lead to a significant increase of the $g$ factor fluctuations, 
including a large probability to find $g$ factors larger than two,
even if the average is well below two.

\begin{figure}
\epsfxsize= 0.75\hsize
\epsffile{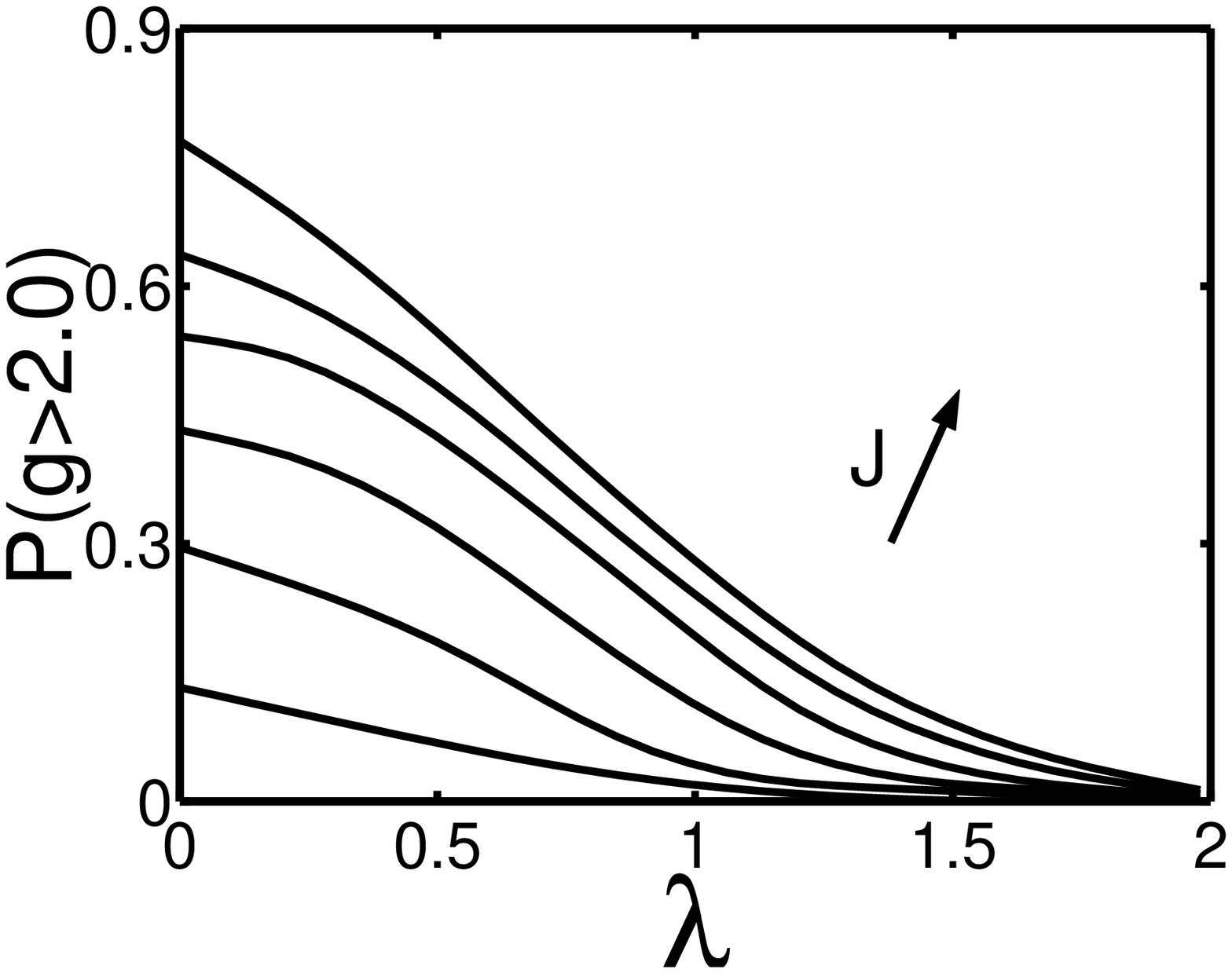}

\epsfxsize= 0.75\hsize
\epsffile{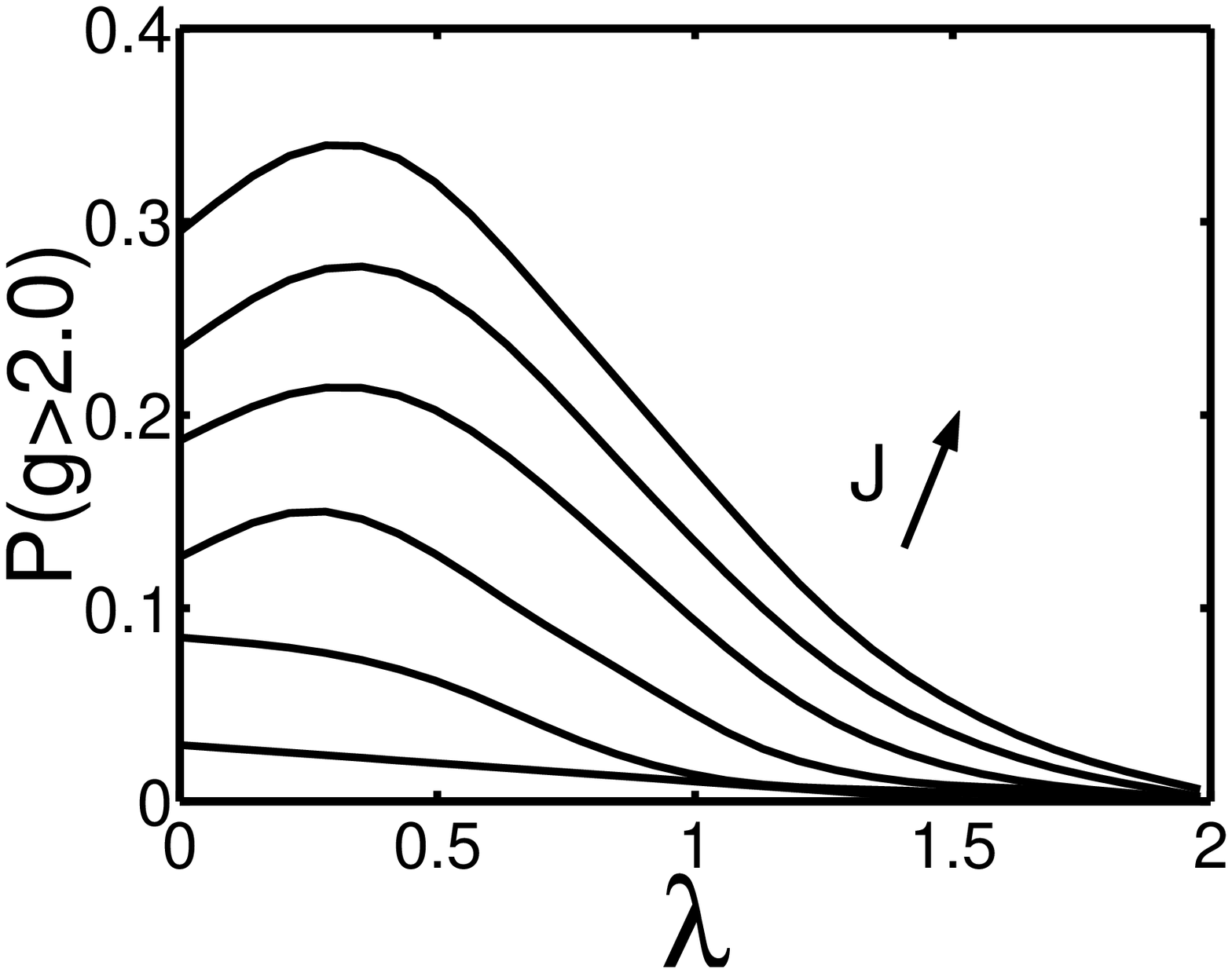}

\caption{
Top: Probability for the $g$ factor of a level to
be larger than $2.0$ for the values of the exchange constant
$J=0.1\div 0.6\delta$ as a function of the spin-orbit scattering
strength $\lambda$
for a random visible (i.e., satisfying the threshold criterion,
see  Sec.\ \ref{sec:threshold}) level.
Bottom: Probability for a random level to be visible and,
at the same time, have $g>2$. 
for the values of the exchange constant
$J=0.1\div 0.6\delta$ as a function of the spin-orbit scattering
strength $\lambda$.
} 
\label{prob_g>2}
\end{figure}

In Fig.~\ref{prob_g>2} we show the probability for a level to
have a $g$-factor larger than two. In an experiment, typically
$g$ factors of 5-10 consecutive levels can be
measured.\cite{kn:petta2001,kn:petta2002} From Fig.\ \ref{prob_g>2}
we then conclude that there is a significant probability that 
one of these $g$ factors is larger than two if $J \agt 0.2 \spacing$.
The bottom panel shows the probability that a level $|k\rangle$ has 
a $g$ factor larger than two {\em and} a weight $\bar w_k > \bar
w_{\rm t}$. As a result of the breakdown of selection rules, this 
probability {\it increases} with increasing spin-orbit scattering
in the region $\lambda\alt 0.5$: States
which have large $g$-factors but small weights for small $\lambda$
become visible for larger values of $\lambda$. The ratio of the
probability shown in the bottom plot of Fig.~\ref{prob_g>2} and
that in the top plot of Fig.~\ref{prob_g>2} is the probability
that a random level can be resolved in the experiment, see Fig.\
\ref{weighted_average_g_factors}

In addition to addressing the full probability distribution of $g$ factors,
we should consider the possibility of correlations between
$g$ factors within one realization. In principle, such correlations
can exist, because, within one realization, all $g$ factors 
correspond to transitions from
the same $N_e$-electron ground state ($N_e$ even). Although the
$N_e$-electron ground state does not affect the values of all
possible $g$ factors for the $(N_e+1)$-particle levels, it does
affect the peak heights, and hence determines which $g$ factors
possibly ``drown'' in the noise.

In order to quantify $g$ factor correlations, we have looked at
the correlation function
\begin{eqnarray}
  C(J,\lambda) &=&
  \left\langle \sum_{k \neq l} \frac{1}{M_{\rm t}^2}
  (g_k - \langle \bar g \rangle)(g_l - \langle \bar g \rangle)
  \right\rangle.
  \label{main_formula_1}
\end{eqnarray}
For the calculation of the correlation function $C(J,\lambda)$
we removed all $g$-factors with normalized weight $\tilde w_k$
below the threshold value $\tilde w_{\rm t} = 0.1$ from the average,
which means that the number levels $M_{\rm t}$ considered in the 
summation becomes dependent on the actual realization.
For the range of exchange interactions $J$ and spin-orbit
scattering rates $\lambda$ we considered, the correlation function
$C(J,\lambda)$ was nonzero, but always smaller than $0.1$. The 
maximum value $C(J,\lambda) \sim 0.1$ was obtained for 
$\lambda \sim 0.5$. Comparing the above difference with
the typical variance of $g$-factors, see Fig.\
\ref{probability_distribution}, we conclude that
correlations between different levels within the same
nanoparticle are not important if the number of $g$ factors measured
in a single metal nanoparticle does not exceed $10$.

\subsection{Probability of nontrivial ground state}

\begin{figure}
\epsfxsize= 0.75\hsize
\epsffile{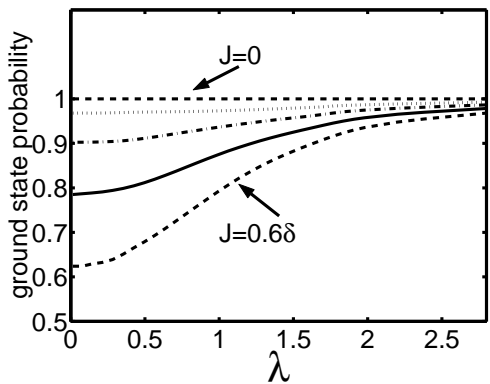}
\caption{Probability to be in the non-interacting ground state
for $J=0.3-0.6\spacing$. The horizontal line $P=1$ corresponds
to the case $J=0$.
} 
\label{probability}
\end{figure}

In Fig.~\ref{probability} we show the probability that the
metal nanoparticle is found in the non-interacting ground state
(Fermi sea), as a function of $\lambda$ and $J$. For $N_e$-electron
states, the probability to find the nanoparticle in the non-interacting
ground state deviates quite significantly from 1 if $J \agt 
0.4$ and $\lambda=0$.\cite{kn:brouwer1999b,kn:baranger2000} 
Upon increasing
$\lambda$, the probability to be in the non-interacting
ground state increases and approaches unity when the spin-orbit
scattering rate exceeds the exchange interaction.

\section{Material dependence}
\label{sec:experiment}

Petta and Ralph have measured the probability
distribution of $g$-factors of Cu, Au, and Ag 
nanoparticles.\cite{kn:petta2001,kn:petta2002}
Theoretical estimates and experimental investigations 
of the exchange interaction in the noble metals show that 
$J/\spacing < 0.1$ for Cu, Au, and 
Ag.\cite{kn:macdonald1982,kn:vier1984}
Hence, the interaction effects in the above materials are very small 
and it is natural that the existing experiments can be explained
quantitatively using theory for the noninteracting ($J=0$) case.
Indeed, both the average and the width of the  $g$ factor probability 
distribution measured in Refs.\ \onlinecite{kn:petta2001,kn:petta2002} 
were found to be in good agreement with the non-interacting
theory of Refs.\
\onlinecite{kn:brouwer2000,kn:matveev2000} using a single fit
parameter, the dimensionless spin-orbit scattering rate $\lambda$. 
The spin-orbit scattering time used in the fits was in 
order-of-magnitude agreement with previous measurements using
weak localization.\cite{kn:bergmann1984} The observation that the
width of the distribution agreed well with theory after $\lambda$
has been chosen to fit the average was considered a success for 
random matrix theory.\cite{foot1}

\begin{figure*}
\epsfxsize=0.9\hsize
\epsffile{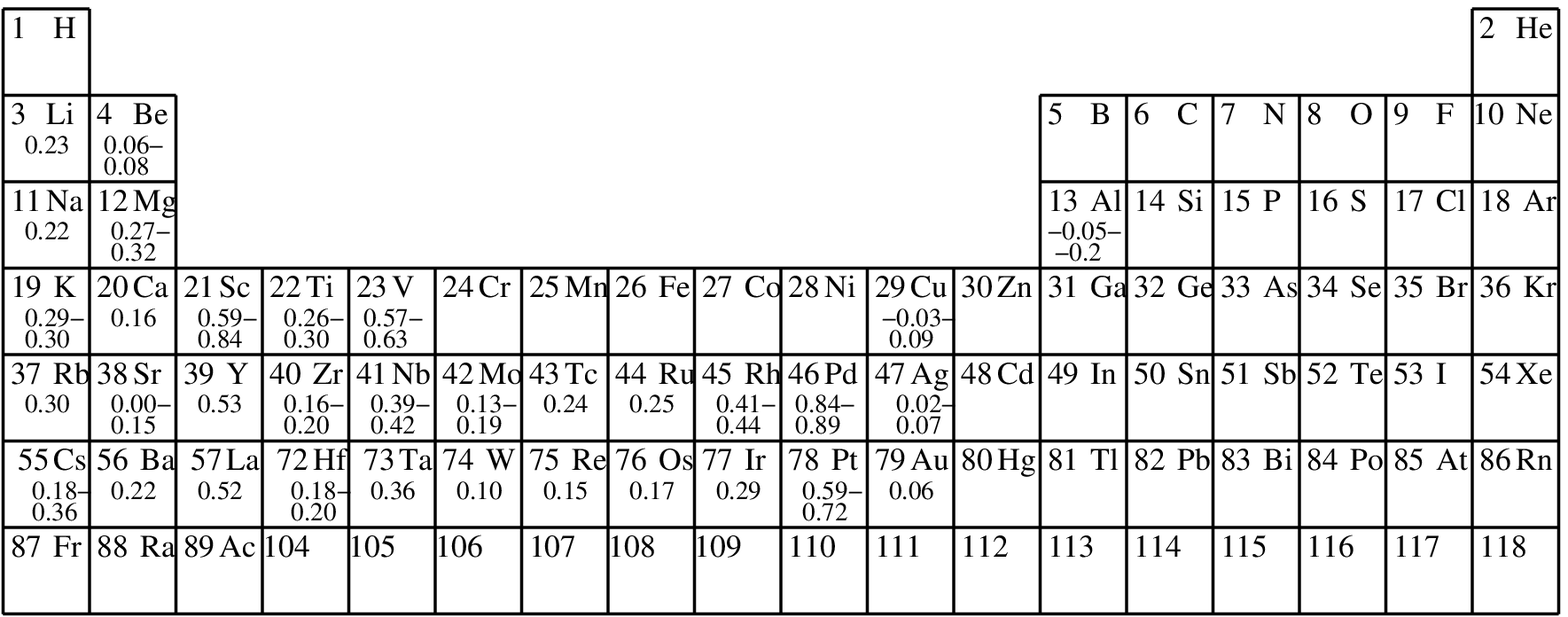}
\caption{Ratio $J/\spacing$ of the exchange interaction constant
and the mean level spacing.
No value is listed for metals from $Z = 24$ to 28 as they are magnetically
ordered as well as all lanthanides ($Z=58\dots 71$) except for
Pm ($Z=61$) for which no value was found in the literature.
Also, for $Z>87$ no data was found in the literature. 
For metals in the right-hand side of the periodic table
(12th column and further) only data on their Pauli susceptibility 
in the liquid form are available, see Ref. \onlinecite{kn:dupree1971}.
For other metals, the data are taken from:
Li, experiment;\cite{kn:vier1984} 
Be, calculation of electronic 
  structure;\cite{kn:wilk1978,kn:janak1977}
Al, experiment\cite{kn:dunifer1977} 
  (note the negative value of the exchange constant);
K, experiments;\cite{kn:knecht1975,kn:randles1972}; 
Ca, Y, Tc, Ba, La, Ta, W, Re, Os, Ir,
  calculation of electronic structure;\cite{kn:sigalas1994}
Sc, calculation of electronic 
  structure;\cite{kn:macdonald1977a,kn:sigalas1994}
Ti, Zr, Hf, calculation of electronic 
  structure;\cite{kn:bakonyi1993,kn:sigalas1994}
V, fit of theory\cite{kn:stenzel1986} and 
  experiment,\cite{kn:hechtfischer1976}, and
  calculation of electronic structure;\cite{kn:janak1977,kn:sigalas1994}
Cu, Ag, experiment \cite{kn:vier1984} and calculation of 
  electronic structure\cite{kn:macdonald1982} (note the negative 
  exchange constant in the experiment for Cu);
Rb, Cs, experiment;\cite{kn:knecht1975}
Sr, Nb, Mo, Rh, calculation of electronic
  structure;\cite{kn:janak1977,kn:sigalas1994}
Pd, fit of theory\cite{kn:stenzel1986} and
  experiment;\cite{kn:manuel1963}
Pt, fit of theory\cite{kn:fradin1975} and 
  experiment,\cite{kn:budworth1960} and calculation of 
  electronic structure;\cite{kn:sigalas1994}
Au, calculation of electronic structure.\cite{kn:macdonald1982}
\label{periodic_table}
}
\end{figure*}

Significant deviations from the non-interacting theory can
be expected for $J/\spacing \gtrsim 0.2$ only. Although this
condition is not satisfied for the noble metals, the exchange
interaction is strong enough to significantly
affect the $g$ factor distribution in most other metals, see 
Fig.\ \ref{periodic_table}, where a list of values of 
$J/\spacing$ reported in the experimental and theoretical literature
is given. The exchange interaction is particularly strong in 
Sc, V, Y, Nb, Rh, and, especially, in Pd (Pd is very close to the 
Stoner instability $J/\delta = 1$).

For the collection of the data shown in Fig.\ \ref{periodic_table},
we used the fact that the ratio $J/\delta$ is related to the Fermi 
liquid parameter $F_0^a$
\be
  J/\spacing = -F_0^a,
\ee
see Ref.\ \onlinecite{kn:oreg2002}. The Fermi liquid parameter
$F_0^a$ appears in the expression for the paramagnetic susceptibility
\be
 \chi = \chi_0 \frac{m^*/m}{1 + F_0^a},
\ee
where $m^*$ is the effective electron mass, including band
structure effects and interaction effects, $m$ is the free
electron mass, and $\chi_0$ is the Pauli susceptibility for free
electrons,\cite{kn:pines1966}
\be
  \chi_0 = \frac{\mu_B^2 m p_F}{\pi^2 \hbar^3}.
\ee
The parameter $1/(1+F_0^a)$ is also known as the
Stoner enhancement parameter.

The dimensionless spin-orbit scattering parameter
$\lambda$ increases with element's nucleus charge $Z$.
In experiments of Petta and Ralph\cite{kn:petta2001} 
strong spin-orbit scattering was found for Au nanoparticles of
a few nm in diameter ($\lambda \sim 0.1$), whereas moderate 
spin-orbit scattering strengths ($\lambda \simeq 1$) were 
found in Cu and Ag nanoparticles of roughly equal size. From
this, we conclude that moderate spin-orbit scattering strengths
can be expected for nanoparticles with a value of $Z$ around those 
for Cu or Ag. From Fig.\ \ref{periodic_table} it can be seen 
that there are quite a few materials for which
this is true and the criterion $J/\delta > 0.2$ is satisfied.

\section{Conclusion}
\label{sec:conclusion}

In this work we investigated the combined influence of 
electron-electron interactions and spin-orbit scattering on the
$g$ factors of metal nanoparticles. In the presence of
electron-electron interactions, $g$ factors must be attributed
to (transitions between) many-electron states, instead of
single-electron states. Many-electron states can have $g$ factors
larger than two, although these cannot be observed by tunneling 
spectroscopy because of
``selection rules'' as long as spin-rotation symmetry is present. 
Spin-orbit scattering breaks spin-rotation symmetry and thus
removes the ``selection rules''. While this leads to a suppression
of the $g$ factors for large spin-orbit scattering rates, we find 
that $g$ factors larger than two occur with significant probability
if the spin-orbit scattering rate $1/\tau_{\rm so}$ is moderate, 
$\tau_{\rm so} \spacing \lesssim 1$, where $\spacing$ is the mean
spacing between single-electron energy levels in the nanoparticle. 
We have studied the $g$ factor distribution quantitatively
using random matrix theory and the universal interaction
Hamiltonian.\cite{kn:kurland2000,kn:aleiner2002} In addition
to a confirmation of the scenario outlined above --- occurrence
of $g$ factors larger than two --- we found that interactions
increase the width of the $g$ factor distribution for 
$\tau_{\rm so} \spacing \lesssim 1$ and 
that the $g$ factors probability distribution function 
is different for transitions to the odd-electron ground state and
odd-electron excited states.
The enhanced fluctuations occurring for moderate spin-orbit
scattering strengths (spin-orbit scattering rate $\gamma_{\rm so}$
comparable to single-electron level spacing $\spacing$) may be
typical of enhanced fluctuations of interaction matrix elements
that are expected to occur at
the breakdown of random matrix theory (dimensionless conductance
$\sim 1$).

There are $\sim 20$ metallic elements
in the periodic table for which the electron-electron interactions
are sufficiently strong that the phenomena described here can
be measured. Existing measurements of $g$ factors in nanoparticles
have been made for Al and the noble metals only; in these metals,
interaction effects are weak. We hope that our findings stimulate 
experiments on other metals.

In our calculations we have omitted the orbital contribution to the
$g$ factors. The orbital contribution arises from the fact that
single-electron wavefunctions are complex if spin-orbit scattering 
is present, instead of real, so that they carry a finite current 
density. For the parameter regime we consider here, $\tau_{\rm so}
\spacing \gtrsim 1$, the single-electron wavefunctions are mostly
real, and we expect the orbital contribution to the $g$ factor to
be smaller than the spin
contribution.\cite{kn:matveev2000,kn:adam2002b}
We expect that taking into account a (small) orbital contribution
causes a slight increase of the average $g$ factor and a further
broadening of the distribution.

\section*{Acknowledgments}
We thank S.~Adam, D.~Huertas-Hernando, 
A.~Kaminski, A.~H.~MacDonald, J.~Petta and D.~C.~Ralph 
for helpful discussions.
This work was supported by the NSF under grant 
no.\ DMR 0086509 and by the Packard foundation.
\begin{appendix}     

\section{$g$ factors in the weak spin-orbit scattering limit}

In this appendix, we calculate $g$ factors to lowest order in the
dimensionless spin-orbit scattering rate $\lambda$. 
For a state that is twofold degenerate in the absence of
spin-orbit scattering, spin-orbit scattering does not affect the 
spin contribution to $g$ factors up to linear order in
$\lambda$.\cite{kn:sone1977} Our calculation addresses the case
of a fourfold degenerate level, and shows that its $g$
factor is affected to zeroth order in $\lambda$.
This calculation
shows explicitly that the limit $\lambda \to 0$ is singular.

\begin{figure}[t]
\epsfxsize= 0.75\hsize
\epsffile{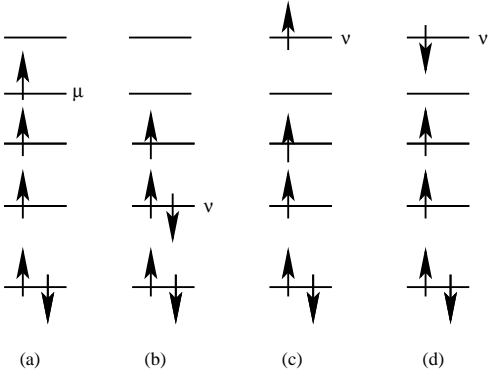}
\caption{ \label{weak_SOS}
An $S=3/2$ state and the three relevant types of 
excited states;
(a)the four-fold degenerate $S=3/2$ ground state; (b) 
two-fold degenerate $S=1/2$ excited state;
(c)four-fold degenerate $S=3/2$ 
excited state; (d) four-fold degenerate $S=1/2$ excited state,
note that this state is entangled
see Eq.~(\ref{s=-1/2})   
} 
\end{figure}

Let us start by writing the spin-orbit Hamiltonian in terms
of the basis of single-electron eigenstates of the Hamiltonian
$H_{\rm GOE}$ without spin-orbit scattering, see Eq.\
(\ref{eq:H0}) above,
\begin{eqnarray}
  H_{\rm so} &=& \frac{i \lambda}{2 \sqrt{N}}
  \sum_{\mu} \left[
  (A_3)_{\mu\nu} (\hat \psi^{\dagger}_{\mu\uparrow}
  \hat \psi^{\vphantom{\dagger}}_{\nu\uparrow} - 
  \hat \psi^{\dagger}_{\mu\downarrow}
  \hat \psi^{\vphantom{\dagger}}_{\nu\downarrow}) 
  \nonumber \right. \\ && \left. \mbox{} +
  (A_1 + i A_2)_{\mu\nu} \hat \psi^{\dagger}_{\mu\uparrow}
  \hat \psi^{\vphantom{\dagger}}_{\nu\downarrow}
  \nonumber \right. \\ && \left. \mbox{} +
  (A_1 - i A_2)_{\mu\nu} \hat \psi^{\dagger}_{\mu\downarrow}
  \hat \psi^{\vphantom{\dagger}}_{\nu\uparrow} \right].
  \label{eq:Hsoapp}
\end{eqnarray}
Here the Greek indices $\mu$ and $\nu$ refer to the eigenvalues
of $H_{\rm GOE}$, and not to the eigenvalues of the total
single-electron Hamiltonian $H_{\rm GOE} + H_{\rm so}$ as in
Sec.\ \ref{sec:theoretical_description}.

In order to study the effect of spin-orbit scattering on the
$g$ factor of a many-electron eigenstate of $H_{\rm GOE}$ with
spin $S=3/2$, one needs to calculate matrix elements between 
the four members of the quadruplet. Labeling the four members
of the quadruplet by the $z$ component of the spin,
$S_z = p - 5/2$, $p=1,2,3,4$, these matrix elements can be
arranged in a $4 \times 4$ matrix $V$ of the form
\be
V = 
\left (
\begin{array}{cccc}
  - a-d   & b        & c          & 0  \\ 
  b^{*}   & - a + d  & 0          & c  \\
  c^{*}   & 0        &  - a + d   & -b \\
  0       & c^{*}    &  - b^{*}   & - a - d 
\end{array}
\right ),
\label{general_form}
\ee
with $a$ and $d$ real numbers and $b$ and $c$ complex numbers. 
The specific form of (\ref{general_form}) follows from 
time-reversal symmetry and guarantees that that the eigenvalues 
of $V$ are double degenerate, in accordance with Kramers' theorem.

One quickly verifies that all matrix elements of $V$ are zero
to first order in $H_{\rm so}$. This is the consequence of the 
fact that the matrices
$A_j$, $j=1,2,3$, in Eq.\ (\ref{eq:Hsoapp}) are antisymmetric,
so that the spin-orbit interaction does not mix the states
with opposite spin belonging to the same energy level. 
In this situation one has to calculate elements of $V$ to 
second order in $H_{\rm so}$.\cite{kn:landau1977} Denoting
the many-electron states with roman indices, the matrix elements
between the many-electron states $n$ and $n^{\prime}$ (both
taken from the same quadruplet) are given by 
\be
  V_{n n^{\prime}} = \sum\limits_{m} 
  \frac{(H_{\rm so})_{nm} (H_{\rm so})_{mn'} }
  {E_{n}^{(0)} -E_{m}^{(0)}}, 
\label{perturbation_theory}
\ee
with $m$ is summed over all many-electron states with $E_{m}
\neq E_{n}$ and $E_{n}^{(0)}$ and $E_m^{(0)}$ the corresponding 
many-electron energies.

The quadruplet state is represented schematically in Fig.\
\ref{weak_SOS}a.
For the calculation of the splitting of a quadruplet, it is 
enough to consider states $m$ of the form indicated in Figs.\
Fig.~\ref{weak_SOS}b, c, and d. These are: a twofold degenerate
$S=1/2$ state (Fig.\ \ref{weak_SOS}b), a fourfold degenerate $S=3/2$ 
state (Fig.\ \ref{weak_SOS}c) and a
fourfold degenerate $S=1/2$ state (Fig.\ \ref{weak_SOS}d).
There exist two variants of the states shown in Fig.\ \ref{weak_SOS}c 
and d, depending on whether an empty single-electron level is
filled above the Fermi level, or a hole is created below the Fermi
level. The former case is shown in the figure.
Since the spin-orbit interaction is a one-particle operator,
only the states of the form shown in Fig.\ \ref{weak_SOS}
which differ by not more than one electron-hole excitation 
are important. 

{\em Virtual excitations to two-fold degenerate $S=1/2$ state.}
In this case, the transition from the state $n$ to $m$ involves a 
transition of an electron from the singly-occupied level $\mu$ to 
the already (singly) occupied level $\nu$, see Fig.\ 
\ref{weak_SOS}a and b.
Representing the members of the spin $S=3/2$ quadruplet as
$|3/2,S_z\rangle$ with $S_z = -3/2$, $-1/2$, $1/2$, $3/2$, and
the members of the spin $S=1/2$ doublet as $|1/2,S_z\rangle$
with $S_z = 1/2$, $-1/2$, we find the following matrix elements
of the spin-orbit Hamiltonian $H_{\rm so}$,
\begin{eqnarray}
  \langle 3/2, +3/2 | H_{\rm so} | 1/2, +1/2 \rangle 
  &=& \frac{i \lambda}{\sqrt{4N}}(A_1 - i A_2)_{\mu \nu},
  \nonumber \\
  \langle 3/2, +3/2 | H_{\rm so} | 1/2, -1/2 \rangle &=& 0,
  \nonumber \\
  \langle 3/2, +1/2 | H_{\rm so} | 1/2, +1/2 \rangle &=& 
  - \frac{i \lambda}{\sqrt{3N}}(A_3)_{\mu\nu}, \nonumber \\
  \langle 3/2, +1/2 | H_{\rm so} | 1/2, -1/2 \rangle &=&
  \frac{i}{\sqrt{12 N}} (A_1-i A_2)_{\mu\nu} \nonumber \\
  \langle 3/2, -1/2 | H_{\rm so} | 1/2, +1/2 \rangle &=& 
  - \frac{i}{\sqrt{12 N}} (A_1+i A_2)_{\mu\nu} \nonumber \\
  \langle 3/2, -1/2 | H_{\rm so} | 1/2, -1/2 \rangle &=& 
  - \frac{i \lambda}{\sqrt{3N}}(A_3)_{\mu\nu}, \nonumber \\
  \langle 3/2, -3/2 | H_{\rm so} | 1/2, +1/2 \rangle &=& 0
  \nonumber \\
  \langle 3/2, -3/2 | H_{\rm so} | 1/2, -1/2 \rangle &=&
  -\frac{i \lambda}{\sqrt{4N}}(A_1 + i A_2)_{\mu \nu}. \nonumber \\
\end{eqnarray}
Further, in this case the energy difference 
\begin{equation}
  E_m^{(0)} - E_n^{(0)} = \varepsilon_{\nu} - \varepsilon_{\mu} + 3 J.
\end{equation}
Substituting these matrix elements and the energy difference 
into Eq.~(\ref{perturbation_theory})
we find the following contributions to the elements of the
matrix $V$ of Eq.\ (\ref{general_form}):
\begin{eqnarray}
  b_{1,\mu\nu} &=& \frac{\lambda^2
    (A_1 - i A_2)_{\mu \nu}(A_3)_{\mu\nu} 
    }{4 N (\varepsilon_{\nu} - \varepsilon_{\mu} + 3 J) \sqrt{3}}
  \nonumber \\
  c_{1,\mu\nu} &=& \frac{\lambda^2 (A_1 - i A_2)_{\mu\nu}^2 }
  {4 N (\varepsilon_{\nu} - \varepsilon_{\mu} + 3 J) \sqrt{3} } \nonumber \\
  d_{1,\mu\nu} &=& 
  \frac{\lambda^2 (A_1)_{\mu\nu}^2 + \lambda^2 (A_2)_{\mu\nu}^2 -
  2 \lambda^2 (A_3)_{\mu\nu}^2}{12 N (\varepsilon_{\nu} - 
  \varepsilon_{\mu} + 3 J)}. \nonumber
\end{eqnarray}
(We have not listed the value of $a$ in the matrix $V$ of
Eq.\ (\ref{general_form}) since this coefficient does
not contribute to the $g$ factor and splitting of the $S=3/2$
quadruplet.)

{\em Virtual excitations to four-fold degenerate $S=3/2$ state.}
These excitations involve a transition of an electron from a singly
occupied level $\mu$ to an unoccupied level $\nu$, see Fig.\
\ref{weak_SOS}, or the transition of an electron from a doubly
occupied level $\mu$ to a singly occupied level $\nu$.
Calculating the various matrix elements as before, we find after
somewhat cumbersome algebra
\begin{eqnarray}
  b_{2,\mu\nu} &=& - \frac{\lambda^2
    (A_1 - i A_2)_{\mu \nu}(A_3)_{\mu\nu} 
    }{6 N (\varepsilon_{\nu} - \varepsilon_{\mu}) \sqrt{3}}
  \nonumber \\
  c_{2,\mu\nu} &=& - \frac{\lambda^2 (A_1 - i A_2)_{\mu\nu}^2 }
  {6 N (\varepsilon_{\nu} - \varepsilon_{\mu}) \sqrt{3} } \nonumber \\
  d_{2,\mu\nu} &=& 
  - \frac{\lambda^2 (A_1)_{\mu\nu}^2 + \lambda^2 (A_2)_{\mu\nu}^2 -
  2 \lambda^2 (A_3)_{\mu\nu}^2}{18 N (\varepsilon_{\nu} -
    \varepsilon_{\mu})}.
 \nonumber
\end{eqnarray}

{\em Virtual excitations to four-fold degenerate $S=1/2$ state.}
As in the previous case, 
these excitations involve a transition of an electron from a singly
occupied level $\mu$ to an unoccupied level $\nu$, see Fig.\
\ref{weak_SOS}, or the transition of an electron from a doubly
occupied level $\mu$ to a singly occupied level $\nu$. The four
$S=1/2$ are labeled by $S_z=\pm 1/2$ and by an additional degeneracy
parameter $q = \pm 1$,
\begin{eqnarray}
  |1/2,+1/2,q\rangle &=& \frac{1}{\sqrt{3}}
  \left (
  |\uparrow    \uparrow    \downarrow \rangle +
  |\uparrow    \downarrow  \uparrow   \rangle e^{ i\frac{2\pi q}{3}} 
  \right. \nonumber \\ && \left. \mbox{} +
  |\downarrow  \uparrow    \uparrow   \rangle e^{-i\frac{2\pi q}{3}}
  \right ), \nonumber \\
  |1/2,-1/2,q\rangle &=& \frac{1}{\sqrt{3}}
  \left (
  |\downarrow  \downarrow  \uparrow   \rangle +
  |\downarrow  \uparrow    \downarrow \rangle e^{ i\frac{2\pi q}{3}}
  \right. \nonumber \\ && \left. \mbox{} +
  |\uparrow    \downarrow  \downarrow \rangle e^{-i\frac{2\pi q}{3}}
  \right ).
\label{s=-1/2}
\end{eqnarray}
Performing the calculations as before, we find
\begin{eqnarray}
  b_{3,\mu\nu} &=& \frac{\lambda^2
    (A_1 - i A_2)_{\mu \nu}(A_3)_{\mu\nu} 
    }{6 N (\varepsilon_{\nu} - \varepsilon_{\mu} + 3 J) \sqrt{3}}
  \nonumber \\
  c_{3,\mu\nu} &=& \frac{\lambda^2 (A_1 - i A_2)_{\mu\nu}^2 }
  {6 N (\varepsilon_{\nu} - \varepsilon_{\mu} + 3 J) \sqrt{3} } \nonumber \\
  d_{3,\mu\nu} &=& 
  \frac{\lambda^2 (A_1)_{\mu\nu}^2 + \lambda^2 (A_2)_{\mu\nu}^2 -
  2 \lambda^2 (A_3)_{\mu\nu}^2}{18 N (\varepsilon_{\nu} -
    \varepsilon_{\mu} + 3 J)}.
 \nonumber
\end{eqnarray}

Denoting the set of doubly occupied single-electron levels by 
``0'', the set of singly-occupied single-electron levels by ``1''
and the set of unoccupied single-electron levels by ``2'', we then
sum over all virtual excitations and find
\begin{eqnarray}
  b &=& \sum_{\mu\neq \nu \in 1} b_{1,\mu\nu}
  + \sum_{\mu \in 1} \sum_{\nu \in 2} (b_{2,\mu\nu} + b_{3,\mu\nu})
  \nonumber \\ && \mbox{}
  + \sum_{\mu \in 0} \sum_{\nu \in 1} (b_{2,\mu\nu} + b_{3,\mu\nu}),
  \nonumber \\
\end{eqnarray}
and similar expressions for the coefficients $c$ and $d$ in Eq.\
(\ref{general_form}).

The matrix (\ref{general_form}) can be diagonalized for all
values of the parameters $a$, $b$, $c$, and $d$, and the 
corresponding $g$-factors can be found exactly. After diagonalization
we find that the quadruplet is split into two doublets with energy
separation
\begin{equation}
  (\Delta E)^2 = 4 d^2 + 4 |b|^2 + 4 |c|^2.
\label{splitting}
\end{equation}
In the $4 \times 4$ matrix notation of Eq.\ (\ref{general_form}),
the Zeeman Hamiltonian reads
\begin{equation}
  H_{\rm Z} =
  \left( \begin{array}{llll} -3 \mu_B H \\
  & - \mu_B H \\
  && \mu_B H \\
  &&& 3 \mu_B H \end{array} \right).
\end{equation}
Lifting the degeneracy of the two doublets by the Zeeman energy,
we find $g$ factors
\begin{equation}
  g = 2 \sqrt{ \frac{3 |b|^2 + (2 d \pm \Delta E)^2}
  {d^2 + |b|^2 + |c|^2}},
\label{g_factor_special}
\end{equation}
where the $\pm$ sign refers to the two doublets. This result
confirms the assertion made earlier, that spin-orbit scattering
affects the $g$ factors of the $S=3/2$ states to zeroth order
in the spin-orbit scattering rate $\lambda$. Of course, for
small $\lambda$ the energy splitting $\Delta E$ is small as
well, and the $g$ factor of Eq.\ (\ref{g_factor_special}) can
be observed for magnetic fields such that $\mu_B H \ll \Delta E$
only, which limits the practical observability of the $g$
factor (\ref{g_factor_special}) for very small spin-orbit
scattering rates $\lambda \ll 1$.

\begin{figure}[t]
\epsfxsize= 0.7\hsize
\hspace{0.1\hsize}
\epsffile{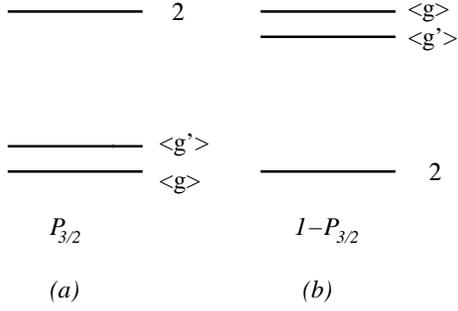}
\caption{ \label{two_different_possibilities}
Two different possibilities: (a) $S=3/2$ state is the ground state
with the probability $P_{3/2}$
(b) $S=1/2$ state is the ground state
with the probability $1-P_{3/2}$. Note that due to the sign
change in the energy denominator, see Eq.~\ref{perturbation_theory}
the average $g$-factors $<g>$ and $<g^{\prime}>$ are inverted.  
} 
\end{figure}

In the remainder of this appendix we investigate Eq.\
(\ref{g_factor_special}) for the special case that one state
is very close to the $S=3/2$ state of interest and virtual
excitations to that state dominate the spin-orbit matrix
$V$ in Eq.\ (\ref{general_form}). One important example is
the case when the ground state has spin $S=3/2$, which is
expected with small probability for $J/\spacing \gtrsim 0.3$, see
Fig.\ \ref{two_different_possibilities}.
Indeed,
the energy difference between the $S=3/2$ state and the 
lowest-lying $S=1/2$ state is 
$\varepsilon_{\lambda+2} - \varepsilon_{\lambda} - 3 J$, 
$\lambda$ being the index of the lowest singly
occupied level in the $S=3/2$ state. Typically, this energy
difference is small, given the small likelihood of it being
positive.

Substituting the general expressions for the coefficients
$b$, $c$, and $d$ into Eq.\ (\ref{g_factor_special}) and
noting that one energy denominator is much smaller than
all others, we find that an $S=3/2$ quadruplet splits into
two doublets with $g$ factors
\begin{eqnarray}
  g &=& 2\sqrt{
  1+
  \frac{3 (A_1^2 + A_2^2)}{A_1^2 + A_2^2 + A_3^2}}, \nonumber \\
  g'  &=&
  \sqrt{48 - 3 g^2}
  \nonumber \\
  &=& 6 \sqrt{\frac{A_3^2}{A_1^2 + A_2^2 + A_3^2}},
\label{simplest_case}
\end{eqnarray}
where we have omitted the indices referring to the levels
$\mu$ and $\nu$ involved as we deal with one excited state
only.  If the quadruplet state is the ground state, $g$ 
corresponds to the lower lying doublet.

In the special case of an $S=3/2$ ground state, Eq.\ (\ref{simplest_case})
shows that the $g$-factor takes values in the interval
$[2,4]$ only. One has $g=2$ 
only if $A_1=A_2=0$. The corresponding eigenstates are
$|3/2,\pm 1/2\rangle$. In the opposite case $A_3=0$, one has
$g=4$ in the ground state and the corresponding eigenstates
are $(\sqrt{3}/{2}) |3/2, \pm 3/2\rangle
- (e^{2 i \phi}/2) |3/2, \mp 1/2\rangle$,
where $e^{- i \phi} = (A_2 + i A_1)/|A_2 + i A_1|$.
In fact, one can find the entire probability distribution of
$g$ in this case, using the fact that $A_1$, $A_2$, and $A_3$
are taken from identical and independent Gaussian distributions,
\begin{equation}
  P(g) = \frac{1}{2 \sqrt{3}}
  \frac{g}{\sqrt{16 - g^2}}. \label{eq:Pg}
\end{equation}
The average $g$-factor for an $S=3/2$ ground state with a 
nearby $S=1/2$ state is then
\begin{eqnarray}
\langle g \rangle & = &
\frac{4}{\sqrt{3}}\left (
\frac{\pi}{2} - \arcsin{\frac{1}{2}} + \frac{\sqrt{3}}{4} 
\right ) \nonumber \\
&\approx & 3.418 .
\end{eqnarray} 
For $J=0.6 \delta$, the probability to find a ground state with
$S=3/2$ is $P_{3/2} \approx 0.38$ (see Ref.\ \onlinecite{kn:oreg2002} or
Fig.\ \ref{probability}). Since the $g$ factor is unaffected to
first order in $\lambda$ if the ground state has spin $1/2$, we
expect the true average ground state $g$ factor to be approximately
equal to
\be
  \langle g_{0} \rangle \approx 3.418 P_{3/2} + 2 (1 - P_{3/2})
  \approx 2.54.
\ee
This value is very close to that found in the numerical simulations
of the ground state $g$-factor $\approx 2.5$, see 
Fig.~\ref{ground_state_g_factor}.

The distribution of $g'$ can be found from Eqs.\
(\ref{simplest_case}) and (\ref{eq:Pg}). One finds the particularly
simple result
\be
  P(g') = \frac{1}{6},\ \ 0 < g' < 6.
\ee
The average $\langle g' \rangle = 3$.
In the limiting cases $g' = 6$ and $g'=0$ the doubly-degenerate
eigenstates 
have the form $| {3/2,\pm 3/2}\rangle $ and $(1/2)|3/2, \pm 3/2\rangle
- (\sqrt{3}/2)e^{2i\phi}|3/2 , \mp 1/2\rangle$, respectively. We
can use these results to calculate the average $g$-factors of the
lowest excited states. For the first excited state
\be
\langle g_1 \rangle = \langle g^{\prime}\rangle P_{3/2} + 
\langle g^{\prime}\rangle
\left (1 - P_{3/2}\right ) = \langle g^{\prime}\rangle = 3 ,
\ee
where we assumed that without spin-orbit scattering the ground
state has spin $S=3/2$ or the the first excited state has spin 
$S=3/2$ and is slightly below an $S=1/2$ doublet, see Fig.\
\ref{two_different_possibilities}. (Hence, we neglect the 
possibilities that the first excited state has spin $S=1/2$ or
that the first excited state has spin $S=3/2$ and is far away
from the next $S=1/2$ state. Our approximation should slightly
overestimate $\langle g_1 \rangle$.)
The simulations give $\langle g_1 \rangle\approx 3.00$, see 
Fig.~\ref{ground_state_g_factor}. With the same
approximations, we find that the average $g$ factor of the
second excited many-electron state is
\be
\langle g_2 \rangle = 2P_{3/2} + \langle g \rangle
\left (1 - P_{3/2}\right )\approx 2.87 .
  \label{eq:st}
\ee
The result of the simulation is $2.78$, in good agreement with the
estimate (\ref{eq:st}).
Note that the above nontrivial distributions
of $g$-factors can be observed in a small 
magnetic field only ($\mu_B H \ll \lambda^2 \delta$). For larger 
fields the $g$-factors are the same as in the
absence of spin-orbit coupling.


\end{appendix}


\begin{thebibliography}{42}
\bibitem[{foo({\natexlab{a}})}]{foot0}
\bibinfo{note}{In principle, there is one more contribution to the interaction
  Hamiltonian. This is a ``pairing interaction'', which, if attractive, causes
  a superconducting instability at sufficiently low temperatures. We consider
  normal metal nanoparticles for which the pairing interaction is small}.
\end{thebibliography}


\end{document}